\documentclass[aps, pra,twocolumn,superscriptaddress,secnumarabic,amssymb,nobibnotes]{revtex4-2}

\usepackage{amsmath,amssymb,graphicx,verbatim}
\usepackage{bbold,color} 
\usepackage[normalem]{ulem}
\usepackage{diagbox}
\usepackage[margin=1in]{geometry}
\newcommand{\comments}[1]{}

\def\({\left(}
\def\){\right)}

\def\ad{a^\dagger}

\def\eq#1{Eq.~(\ref{eq:#1})}

\def\eqs#1#2{Eqs.~(\ref{eq:#1}-\ref{eq:#2})} 
 
\def\fig#1{Fig.~\ref{fig:#1}}
\def\Fig#1{Figure~\ref{fig:#1}}
\def\bra#1{\left\langle\,#1\,\right|} 
\def\ket#1{\left|\,#1\,\right\rangle}
\newcommand\sect[1]{Section~\ref{sec:#1}}
\newcommand\app[1]{Appendix~\ref{app:#1}}
\newcommand\tb[1]{Table~\ref{tab:#1}}

\begin{document}

\title{Fock state interferometry for quantum enhanced phase discrimination}
\author{Reihaneh Shahrokhshahi}
\affiliation{University of Virginia, Dept. of Physics, 382 McCormick Rd., Charlottesville, VA 22904-4714, USA}
\author{Saikat Guha}
\affiliation{College of Optical Sciences, University of Arizona, Tucson, AZ 85721, USA}
\author{Olivier Pfister}
\email{opfister@virginia.edu}
\affiliation{University of Virginia, Dept. of Physics, 382 McCormick Rd., Charlottesville, VA 22904-4714, USA}
\date{\today}
\begin{abstract}
We study Fock state interferometry, consisting of a Mach-Zehnder Interferometer with two Fock state inputs and photon-number-resolved detection at the two outputs. We show that it allows discrimination of a discrete number of apriori-known optical phase shifts with an error probability lower than what is feasible with classical techniques under a mean photon number constraint. We compare its performance with the optimal quantum probe for $M$-ary phase discrimination, which unlike our probe, is difficult to prepare. Our technique further allows discriminating a null phase shift from an increasingly small one at zero probability of error under ideal conditions, a feature impossible to attain using classical probe light. Finally, we describe one application to quantum reading with binary phase-encoded memory pixels.
\end{abstract}
\maketitle

\section{Introduction}
Quantum mechanics dictates the fundamental limit of phase estimation. A typical interferometer, such as the Michelson or Mach-Zehnder interferometer, probes a phase shift by splitting an input field into two mutually coherent fields, each probing a different path. The fields are then recombined and the resulting total intensity yields interference fringes which inform on the phase-difference between the two paths. The quantum physics of interferometry hinges on the initial splitting of the input field: by unitarity, the two split outputs necessarily call for {\em two} inputs, in of which is in a vacuum state. As shown by Carl Caves~\cite{Caves1980}, {\color{red} it} the yields the classical --- a.k.a.\ beam-splitter shot-noise --- limit 
\begin{equation}\label{eq:cl}
\Delta \theta_{cl} \sim {\langle N \rangle}^{-\frac12},
\end{equation}
where $ \theta$ is the  phase difference between the arms of an interferometer and $\langle N \rangle$ is the total average number of photons in the interferometer. The classical limit, however, doesn't give the ultimate phase precision, which is fixed by the Heisenberg number-phase Heisenberg inequality~\cite{JMLL} and bounded by the Heisenberg limit (HL)
\begin{equation}\label{eq:hl}
\Delta \theta_{Hl} \sim {\langle N \rangle}^{-1}.
\end{equation}
In order to reach the HL, the vacuum input must be replaced by a nonclassical state such as a squeezed state~\cite{Caves1981}. Other possibilities were proposed~\cite{Yurke1986a}, in particular the use of correlated Fock states~\cite{Holland1993} or superpositions or statistical mixtures thereof~\cite{Kim1998}. Many theoretical studies of nonclassical inputs yielding the HL have been conducted~\cite{Luis2000}, in particular applying information theory to quantum physics~\cite{Helstrom_1976}, which has led to the definition of quantum metrology~\cite{Giovani:2006_Qmetrology}. In addition, a recent study has shown that ensuring phase estimation at the HL requires, in general, losses to be bounded by ${\langle N \rangle}^{-1}$~\cite{Escher2011}.

\begin{figure}[h!]
\includegraphics[width=\columnwidth]{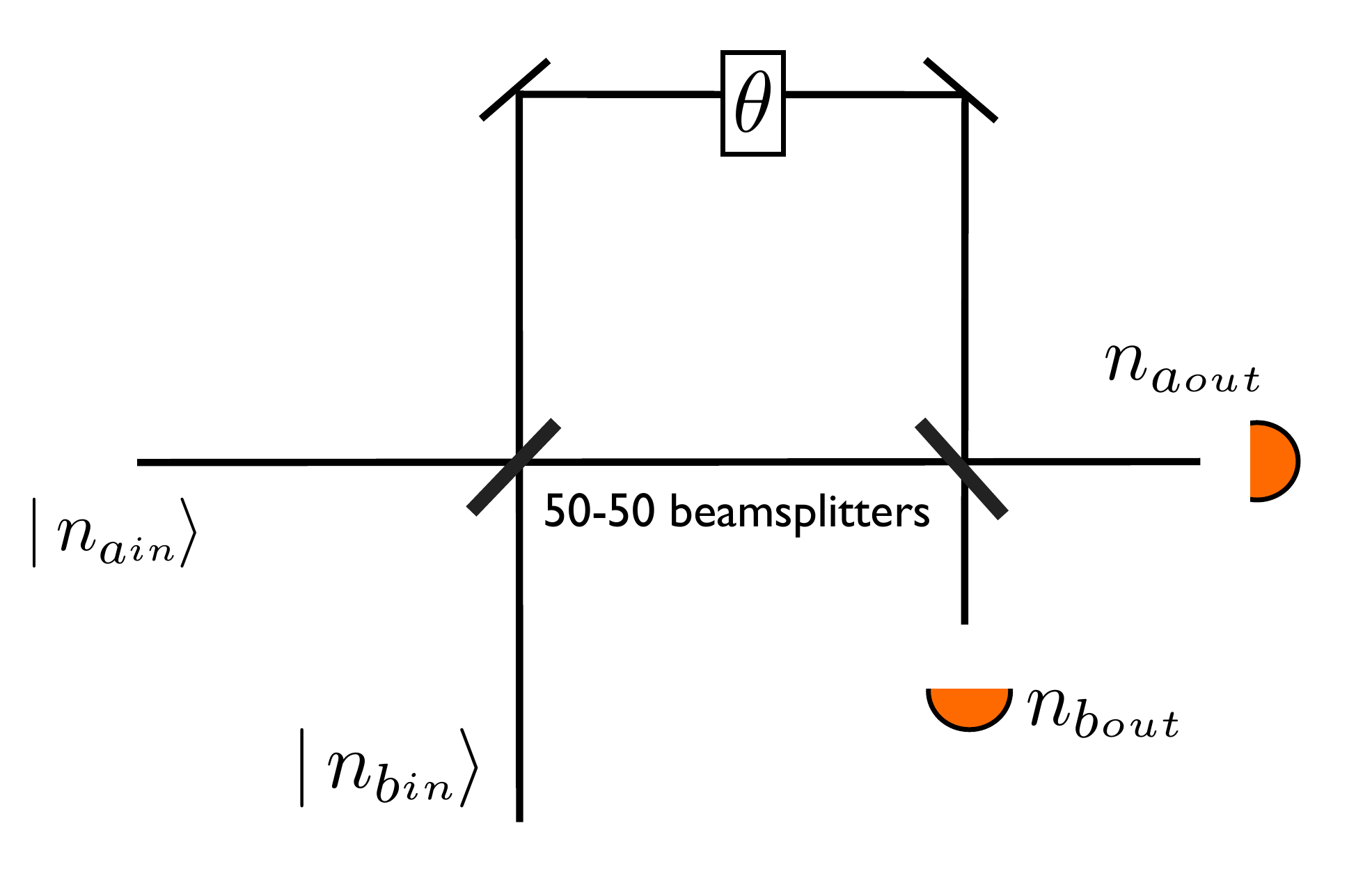}
\caption{Phase discrimination by Fock-state interferometry. The Fock state $\ket{n_a}_{a}\ket{n_b}_{b}$ is input into a Mach-Zehnder interferometer of phase difference $\theta$ and the interferometer's  output is measured by  photon-number-resolving detectors.}
\label{fig:Imaging-setup}
\end{figure}

In this paper, we investigate the properties of the highly nonclassical, yet conceptually simple interferometry which we call Fock-state interferometry (FSI), as depicted in \fig{Imaging-setup}: in all of the paper, FSI consists in a Mach-Zehnder Interferometer (MZI) with Fock state input and photon-number-resolved detectors (PNRDs).
One should note that FSI was recently implemented experimentally, and the effect of photon loss was studied~\cite{Thekkadath2020}. FSI was also studied theoretically with multimode inputs~\cite{Perarnau_Llobet_2020}. Here, we focus on FSI not for phase estimation as much as for phase discrimination.

It may be useful to give a simple insight on the expected impact of such a purely corpuscular interferometer input on purely undularory interferometric performance. It is simple to show, as we do below, that FSI benefits general phase estimation only when the input state satisfies $n_{a}=n_{b}$~\cite{Holland1993}. We outline a basic derivation, using the Schwinger spin representation of two boson fields~\cite{Schwinger1965}, which we'll use in all of the paper and which is presented, along with its application to quantum interferometry, in \app{sr}. The Schwinger spin formalism allows one to use the convenient formalism of quantum angular momentum and SO(3) rotation matrices to perform quantum optics calculations. 

To begin with, the expectation value  of the output photon number difference is 
\begin{align}
\langle N_a-N_b \rangle_\text{out}
&=\langle J_{z} \rangle_\text{out} \\
&= 2m\cos\theta\\
&=(n_{a}-n_{b})\cos\theta
\end{align}
which constitutes the quantum expression of an interference fringe. 

Note that, when $n_{a}=n_{b}\Leftrightarrow m=0$, the output photon number difference is zero for all values of $\theta$ hence no direct fringe is present and other methods are required to access phase information. However, these methods yield Heisenberg-limited performance~\cite{Holland1993,Kim1998}. 

When $n_{a}\neq n_{b}\Leftrightarrow m\neq0$, a direct interference signal is present but we show that direct measurement performance of FSI is, in fact, at the classical limit.
First, we evaluate the quantum standard deviation of the output photon-number difference, using $j=(n_{a}+n_{b})/2$ and $m=(n_{a}-n_{b})/2$ for input state $\ket{j\,m}_{z}=\ket{n_{a}}_{a}\ket{n_{b}}_{b}$, 
\begin{align}
\Delta J_{z}^\text{out} &= |\sin\theta|\,\{2[j(j+1)-m^{2}]\}^{\frac12} 
\end{align}
from which we obtain the theoretical phase error (whose expression is invalid for $n_{a}=n_{b}$)
\begin{align}
\Delta\theta &= \frac{\Delta J_{z}^\text{out}}{\left|\frac{\partial\langle J_{z}\rangle_\text{out}}{\partial\theta}\right|} \\
\Delta\theta &= \left[\frac{j(j+1)}{2 m^{2}}-\frac12\right]^{\frac12},
\end{align}
which is only valid for $m\neq0$, i.e., $n_{a}\neq n_{b}$. Since the total photon number $j$ is a constant here, it is easy to see that the minimum error can only be obtained by maximizing $m^{2}$, i.e., for $m=\pm j$ ($n_{b,a}=0$ respectively), in which case we get
\begin{align}
\Delta\theta_\text{min} &= (2j)^{-\frac12} 
\end{align}
which is the classical limit, \eq{cl}, as was first shown by Caves~\cite{Caves1980}.

We show in this paper that FSI can yield enhanced interferometry performance and break the classical limit in cases less general than the estimation of an unknown phase, namely the discrimination of two or more predetermined phase shifts. The rationales for this study are multiple: {\em (i)}, the theoretical simplicity of the experimental setup; {\em (ii)}, its translation into sophisticated experimental concepts that are nonetheless coming of age, i.e., on-demand photon sources 
and photon-number-resolving detectors; {\em (iii)}, the availability of concrete applications of phase discrimination, such as the quantum reading of classical optical memories.

This paper is organized as follows, in \sect{opd} we  study the problem of discriminating the finite number $M \geqslant 2$  of optical phase shifts for M-ary Phase Shift Keying (MPSK)~\cite{Agrawal_2010_opt_commmunication} and show that our model can discriminate between two or three phases with respectively  zero and near zero error probability using only few photons, and outperforms  phase discrimination schemes using coherent states with heterodyne and homodyne receivers. In \sect{qr}, we study the quantum optical reading of digital memory and show it outperforms its classical counterpart. We then conclude.

\section{Phase discrimination}\label{sec:opd}

\subsection{Introduction}

In optical communication terms, the ability to discriminate between $M$ optical phase shifts  can be beneficial to MPSK,  a digital modulation scheme that conveys $M$ messages by modulating the optical phase of a probe signal. Here we show that using FSI we can accurately discriminate between $M=2,3$ optical phase shifts, using a single photon count. Moreover, we  show that  FSI allows discrimination of the information encoded in increasingly smaller phase shifts. 

For simplicity, we first explain the concept with $M=2$ phases and then generalize it to higher values of $M$. We consider an unknown  fixed  phase $\theta$ of MZI, which can take one of $M=2$ values, denoted $\theta_1$and $\theta_2$  and we define the estimated phase $\hat{\theta}$. Four different scenarios  can occur during the phase discrimination process. If the initial phase $\theta=\theta_{1,2}$ and the estimated phase $\hat{\theta}=\theta_{1,2}$, respectively, then we have success. Else $\hat{\theta}=\theta_{2,1}$, and we have an error.  A natural criterion to measure interferometer performance in the phase discrimination problem will then be the error probability, $P_e$, which we'll define later. 

We consider a MZI with a $\ket{j\,\mu}$ input and whose phase $\theta$ can be either of two predetermined values  $\theta_{1,2}$. We then perform a {\em single} $J_z$ measurement of the photon number difference at the output ports, of result $\mu'$, and make a decision about the  phase shift based on  maximum likelihood algorithm: knowing the probability distribution $P(\mu',\mu|\theta)$ of the interferometer (\tb{sr}), 
\begin{table}[t]
\caption{Probability distribution $P(\mu',\mu|\theta)$. The possible measurement outcomes are denoted by $\mu'$ (columns) and possible phases by $\theta_{1,2}$ (rows). Each element of this array is  the probability of measuring $\mu'\in[-j,j]$, given phase $\theta$.  
}
\[
\begin{tabular}{|c|c|c|c|c|c|}
\hline
\backslashbox{$\theta$}{$\mu'$} & $-j$&...&$m$&...&$j$\\
\hline
&&&&&\\
$\theta_{1}$& $P(-j,\mu|\theta_{1})$&...&$P(m,\mu|\theta_{1})$&...&$P(j,\mu|\theta_{1})$\\
&&&&&\\
\hline
&&&&&\\
$\theta_{2}$& $P(-j,\mu|\theta_{2})$&...&$P(m,\mu|\theta_{2})$&...&$P(j,\mu|\theta_{2})$\\
&&&&&\\
\hline
\end{tabular}
\]
\label{tab:sr}
\end{table}
we compare both cases $\theta=\theta_1$ and $\theta=\theta_2$ for a given measurement outcome and assign the estimated phase shift $\hat{\theta}$ to the phase which is more likely to result in this specific outcome $\mu'$. The algorithm is thus
\[ 
\begin{array}{lc}
\text{if } P(\mu',\mu| \theta_{1}) \geqslant P(\mu',\mu| \theta_{2}) & \\
\text{then}   \left\{\begin{array}{c} P(\hat\theta=\theta_{1}|\theta=\theta_{1})=P(\mu',\mu| \theta_{1})\text{\quad --- success} \\ 
P(\hat\theta=\theta_{1}|\theta=\theta_{2})=P(\mu',\mu| \theta_{2}) \text{\quad --- failure}\end{array}\right.\\
\text{else}   \left\{\begin{array}{c} P(\hat\theta=\theta_{2}|\theta=\theta_{2})=P(\mu',\mu| \theta_{2})\text{\quad --- success} \\ 
P(\hat\theta=\theta_{2}|\theta=\theta_{1})=P(\mu',\mu| \theta_{1}) \text{\quad --- failure}\end{array}\right.
\end{array}
\]
For this procedure to be error free, one would need:
\begin{align}
P(\hat\theta=\theta_{1,2}|\theta=\theta_{2,1})=0.
\end{align}
Of course, this is not the case in general, and the average error probability is given by:
\begin{align}
P_e=\sum_{i,j\neq i}P(\theta_i)P(\hat{\theta}=\theta_j| \theta=\theta_i).
\label{eq:Pe}
\end{align}
However, we now show that, for a judicious choice of phases $\theta_{1,2}$ and input $\mu$,  phase discrimination with an FSI may perform better than classical methods such as a coherent state probe and homodyne or heterodyne detection.

\subsection{Binary phase discrimination}

The analytic expressions of probabilities are given by rotation matrix elements in the Schwinger representation, \eq{dmatrix}, \app{sr}. Without loss of generality, we may elect to set $\theta_{1}=0$ as this entails 
\begin{align}
 P(\mu',\mu|0) = d_{\mu',\mu}^j(0)^{2} = \delta_{\mu',\mu}
\end{align}
and simplifies the situation. The problem will then reduce to discriminating $\theta_{2}=\theta$ against $\theta_{1}=0$. Note that this is still different from general phase estimation --- again classically-limited for a FSI --- as we'll restrict $\theta$ to the values that will allow optimized performance. 

Using information theory --- see \app{it} for a brief review --- we will evaluate two important figures of merit of phase discrimination: first, the error probability, \eq{Pe}, and also the mutual information $I(\theta;\hat\theta)$ between $\hat\theta$ and $\theta$, \eq{MI_reverse}.

\subsubsection{Influence of the total photon number}

We study the behavior of FSI for input states with different total photon number $n_{a}+n_{b}=2j$. We calculate the error probability, and also the mutual Information, for all $\theta$.  We then determine optimal phases for each input state, based on the performance. 

\begin{figure}[htbp]
\begin{center}
\includegraphics[width=.9\columnwidth]{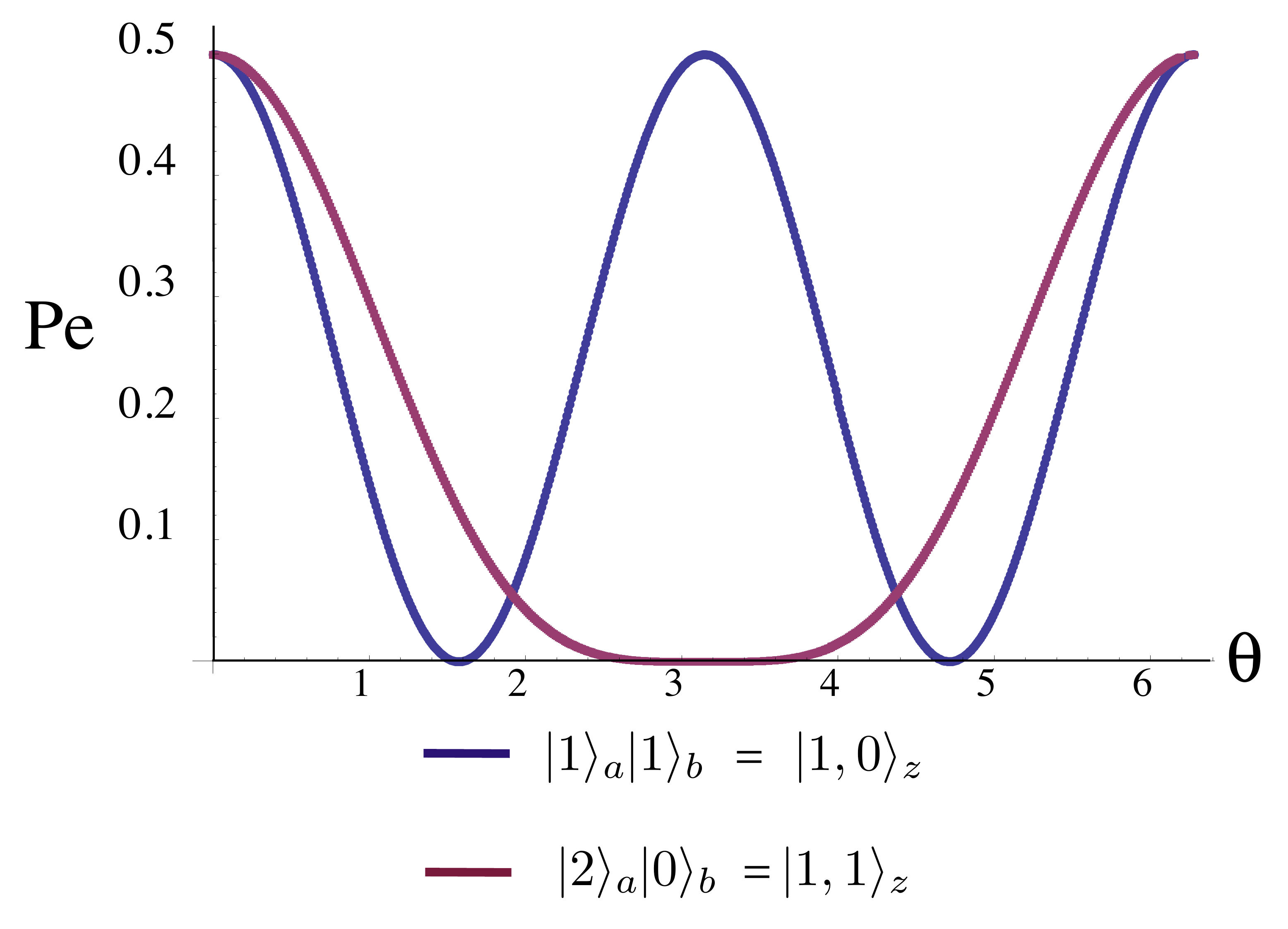}
\caption{Error probability $P_e(\theta)$ for binary phase discrimination with a total photon number $2j=2$.}
\label{fig:Pe(p)j1}
\end{center}
\end{figure}

We first consider binary ($M=2$) phase discrimination for the two-photon inputs $\ket1_{a}\ket1_{b}$ and $\ket2_{a}\ket0_{b}$, which are the respective Schwinger spin states $\ket{1\,0}_{z}$ and $\ket{1\,1}_{z}$. \Fig{Pe(p)j1} displays the error probability versus the phase.
As can be seen on the figure, both inputs yield zero error probability. However, the smallest phase $\theta$ such that $P_{e}(\theta)=0$ is $\pi$ for a ``classical'' FSI (vacuum input, i.e., $m=\pm j$) and $\pi/2$ for an $m=0$ input. An immediate question is therefore whether this trend continues and larger photon numbers yield discrimination of smaller phases.

\begin{figure*}[htbp]
\begin{center}
\includegraphics[width=.9\textwidth]{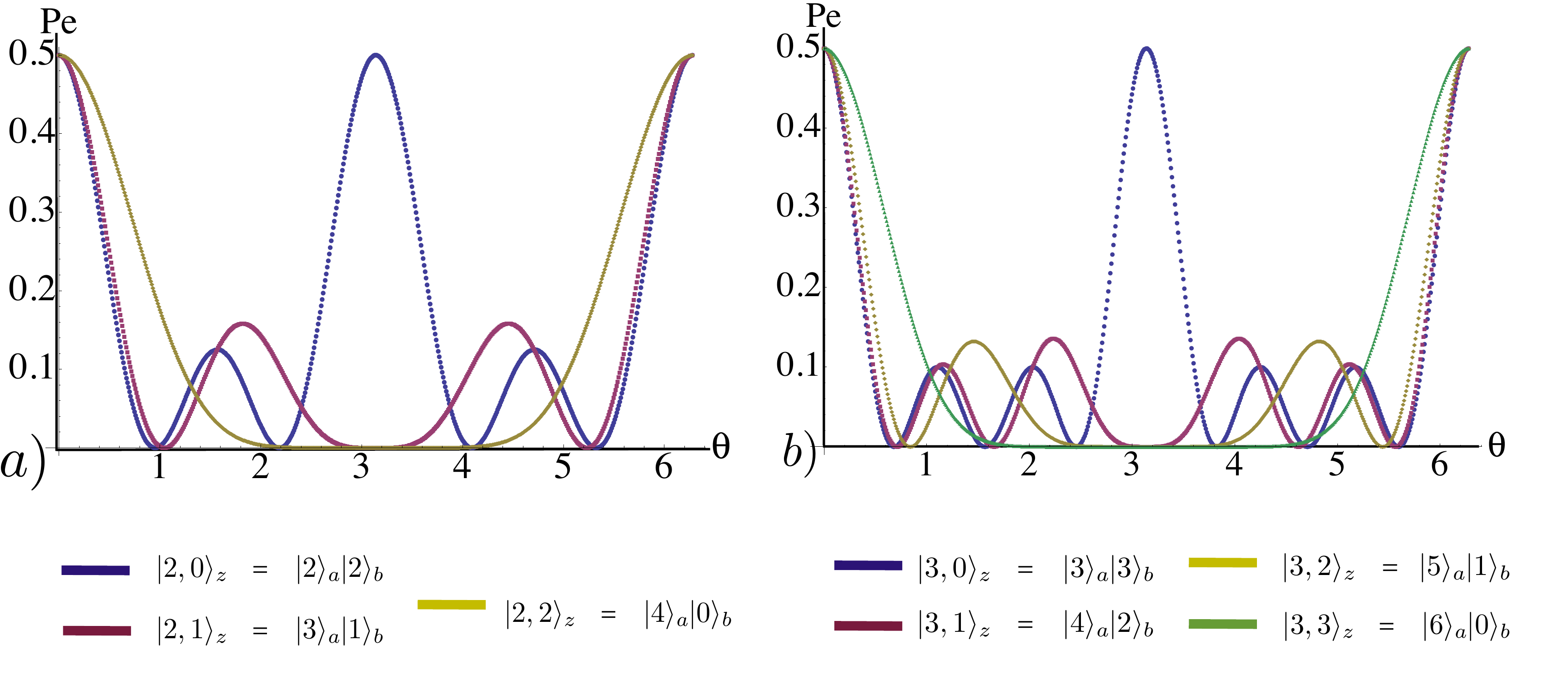}
\caption{Error probability $P_e(\theta)$ for binary phase discrimination with a total photon number of a) $2j=4$, b) $2j=6$.}
\label{fig:Pe(p)j23}
\end{center}
\end{figure*}

The 6-photon ($j=3$) and 8-photon ($j=4$) cases are displayed in \fig{Pe(p)j23}(a) and \fig{Pe(p)j23}(b), respectively.
These situations are richer because more photon number partitions are possible at the interferometer's input. Two trends are clearly visible: {\em (i)},  from \fig{Pe(p)j23}(a) and \fig{Pe(p)j23}(b) individually, i.e., for a given $j$, the first zero of $P_{e}(\theta)$, or ``smallest optimum phase,'' decreases as $|m|$ decreases and, {\em (ii)}, from \fig{Pe(p)j23}(a) and \fig{Pe(p)j23}(b) together, the smallest optimum phase decreases as $j$ increases.

\begin{figure}[b]
\begin{center}
\includegraphics[width=\columnwidth]{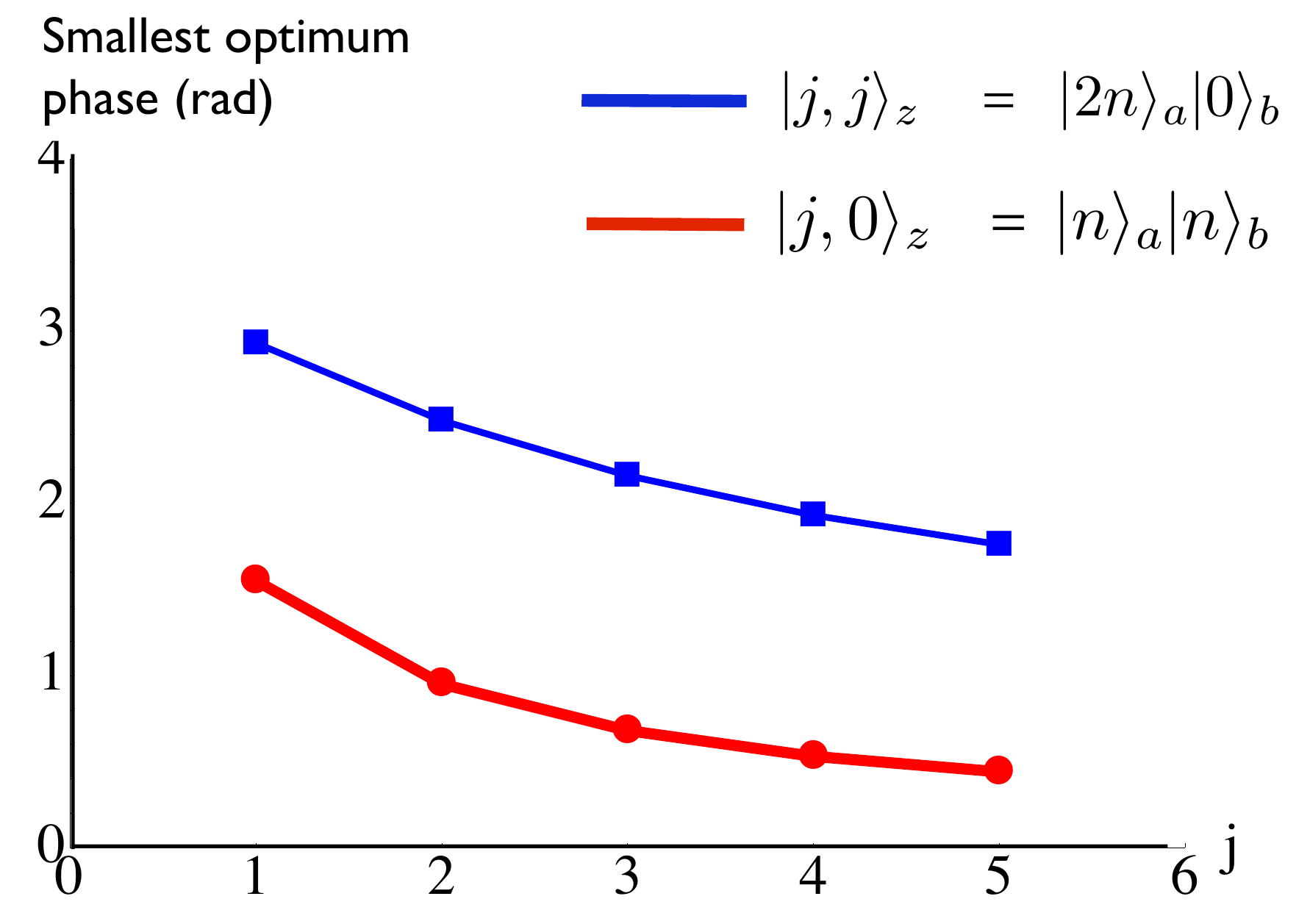}
\caption{Smallest optimum phase vs $j$, The optimum $\theta$ is smallest for twin-Fock state input $\ket{n}_a\ket{n}_b$, $\forall n$.}
\label{fig:opt_p_Allj}
\end{center}
\end{figure}

In order to confirm the latter, we computed the smallest optimum phase $\theta_{\text{sop}}$ for larger photon numbers  in the $m=0$ and $m=j$ cases, results are plotted in \fig{opt_p_Allj}.
For the values up to $2j=10$ computed in \fig{opt_p_Allj}, a fit for the balanced Fock state with $m=0$ (red plot) yields 
\begin{align}
\theta_{\text{sop}} &= 2.67(6)\times (2j)^{-0.77(2)},
\label{eq:sop}
\end{align}
where the parenthetic numbers $(6)$ and $(2)$ represent fit uncertainty. While we do not yet have any certainty about the theoretical form of the dependence of $\theta_{\text{sop}}$ on $j$, these results clearly hint at the ability for ``twin'' Fock states $\ket n_{a}\ket n_{b}=\ket{j=n\ m=0}_{z}$ to allow the discrimination of phase shifts $<j^{-3/4}$, which beats the classical limit $j^{-1/2}$. This is, indeed, proven by our calculations up to $2j=10$.

This result is reminiscent of Holland-Burnett interferometry,~\cite{Holland1993} which employs the same input state. However, the measurement processing is entirely different as it uses maximum likelihood, as opposed to Bayesian reconstruction.
 
We know turn to another form of comparison of FSI phase discrimination with conventional methods, by comparing it with the coherent-state-based homodyne detection and Dolinar receivers, as well as with the use of phase eigenstates.
 
\subsubsection{Comparison with coherent state input methods and with phase eigenstates}

Various other states can be used as interferometer input. Coherent states, in particular, present the advantage of being readily accessible from the output of a well stabilized laser. 

In binary phase-shift-keyed encoding using coherent states, the phase-modulated coherent states are given as $\ket{\alpha}$ and $\ket{-\alpha}$, i.e., two fields of opposite phase. In the low photon number regime, these states have  small amplitudes $(|\alpha|=\sqrt{j})$ and are largely overlapping (due to nonorthogonality of coherent states), so the ability to successfully distinguish these states at the receiver is limited, ultimately by the Helstrom bound~\cite{Helstrom_1976}. Various detection schemes can approach this bound for coherent state discrimination. We consider two of them: the homodyne receiver and the Dolinar receiver. 

The former is the simplest possible receiver relying on Gaussian (Wigner function) operations~\cite{Braunstein_2005_Homodyne_det}.  
The error probability for coherent state discrimination using a homodyne receiver is given by~\cite{Oliviars_2004_homodyne_error}:
\begin{align}
P_e(\alpha)=\frac{1}{2}\left[1-\text{erf}\left(\frac{|\alpha|}{2}\right)\right].
\label{eq:PeHomo}
\end{align}

The latter is an adaptive measurement scheme, proposed by Dolinar in 1973, and which reaches the Helstrom bound for discriminating between two pure coherent states~\cite{Dolinar1973}. The Dolinar receiver is based on a combination of photon counting and real-time feedback control. The minimum error probability using a Dolinar receiver is
\begin{align}
P_e^\text{min}=\frac{1}{2}(1-\sqrt{1-e^{-4 |\alpha|^2}}),
\label{eq:PeDolinar}
\end{align}
which is the lowest possible error in distinguishing between two pure coherent states.  Therefore, one reaches zero error probability only asymptotically, using high-amplitude coherent states.


Finally, we mention a theoretical proposal, which doesn't rely on coherent states, by Nair {\em et al.}~\cite{saikat:2012_symphasedisc} for the quantum state inside (between the two beamsplitters of) the MZI, rather than at its input. When $j \geqslant ({m-1})/{2}$, there always exists an optimum quantum state which can resolve $m$ optical phases. This quantum state is Pegg and Barnett's eigenstate of the optical phase operator~\cite{Pegg1989}  
\begin{align}
\ket{\varphi}=\frac{1}{\sqrt m}\sum_{n=0}^{m-1} e^{-i n \varphi} \ket{n}
\label{eq:phase_state}
\end{align}
taken at $\varphi=0$
\begin{align}
\ket{\varphi=0}=\frac{1}{\sqrt m}\sum_{n=0}^{m-1}  \ket{n}
\label{eq:phase_state}
\end{align}
This state is intended to probe the phase shift in the ``signal'' arm of the MZI, which it can do at the Heisenberg limit. 
For $m=2$ ($j\leqslant1$), the minimum error probability is
\begin{align}
P_e=\frac{1}{2}-\sqrt{j(1-j)}.
\end{align}

\begin{figure}[htbp]
\begin{center}
\includegraphics[width=\columnwidth]{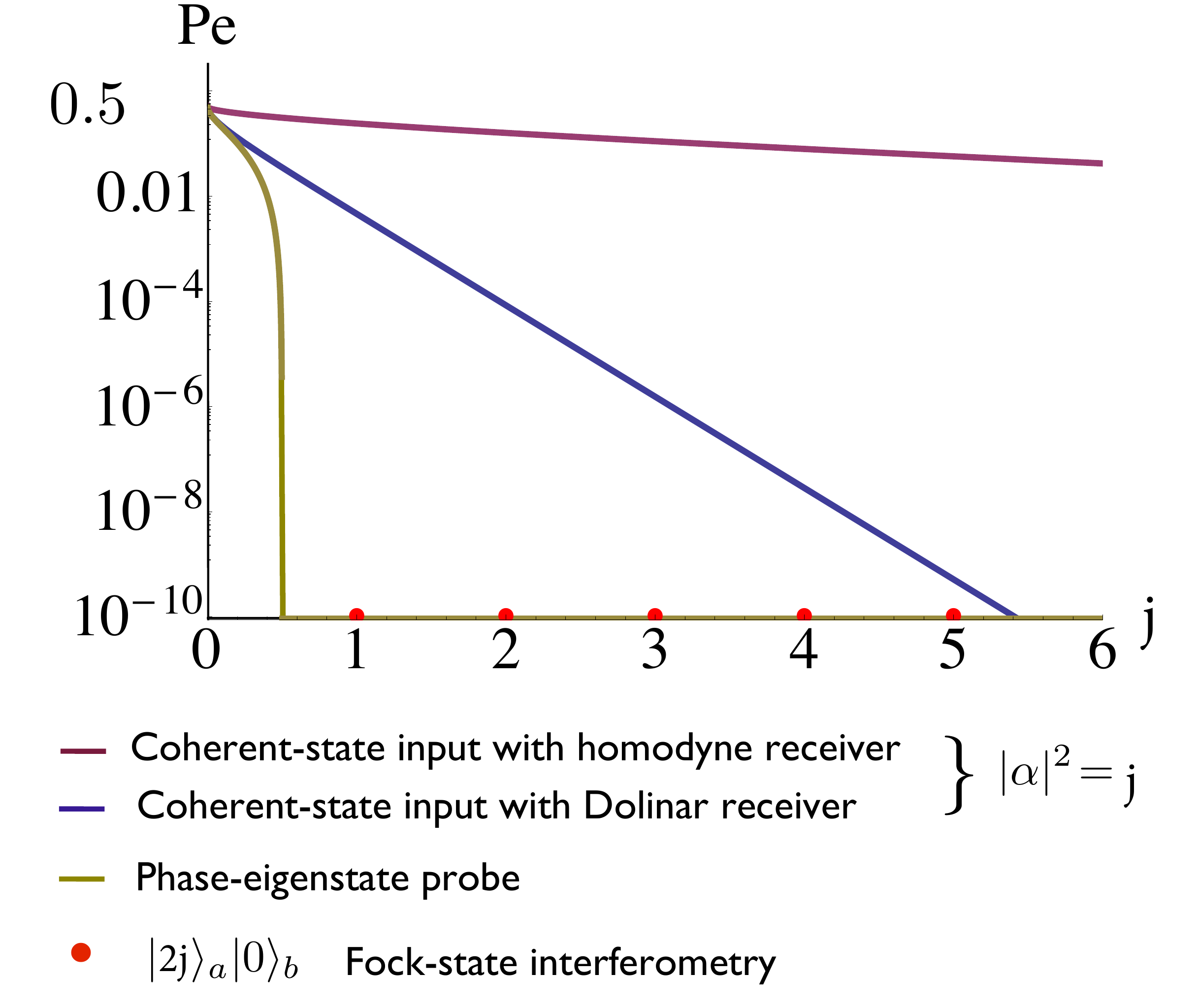}
\caption{Error probability for discriminating between $0$ and $\pi$ radians, versus $j$ (where $2j$ is the total photon number).}
\label{fig:Pe_N__classical}
\end{center}
\end{figure}

We now turn to the comparison of the error probabilities for binary phase discrimination (one of the phases being set to zero) of Fock-state interferometry, coherent-state homodyne detection, the Dolinar receiver, and phase-eigenstate probe. The respective error probabilities for discriminating between $0$ and $\pi$ radians are plotted in \fig{Pe_N__classical}.
As can be seen on the figure, FSI's performance is remarkable as it matches that of the optimal phase-eigenstate probe in this ($0$,$\pi$) case, and, like the phase-eigenstate probe, yields a clear advantage over the Dolinar receiver for small photon numbers. This is also of interest as FSI, unlike the phase-eigenstate probe, has a clear experimental implementation.

In addition, we recall that FSI offers the additional capability of resolving decreasing phase shifts as the total photon number increases (\fig{opt_p_Allj}), as per Eq.~\eqref{eq:sop}.

\subsection{Ternary phase discrimination}

\begin{figure}[h!]
\begin{center}
\includegraphics[width=1.15\columnwidth]{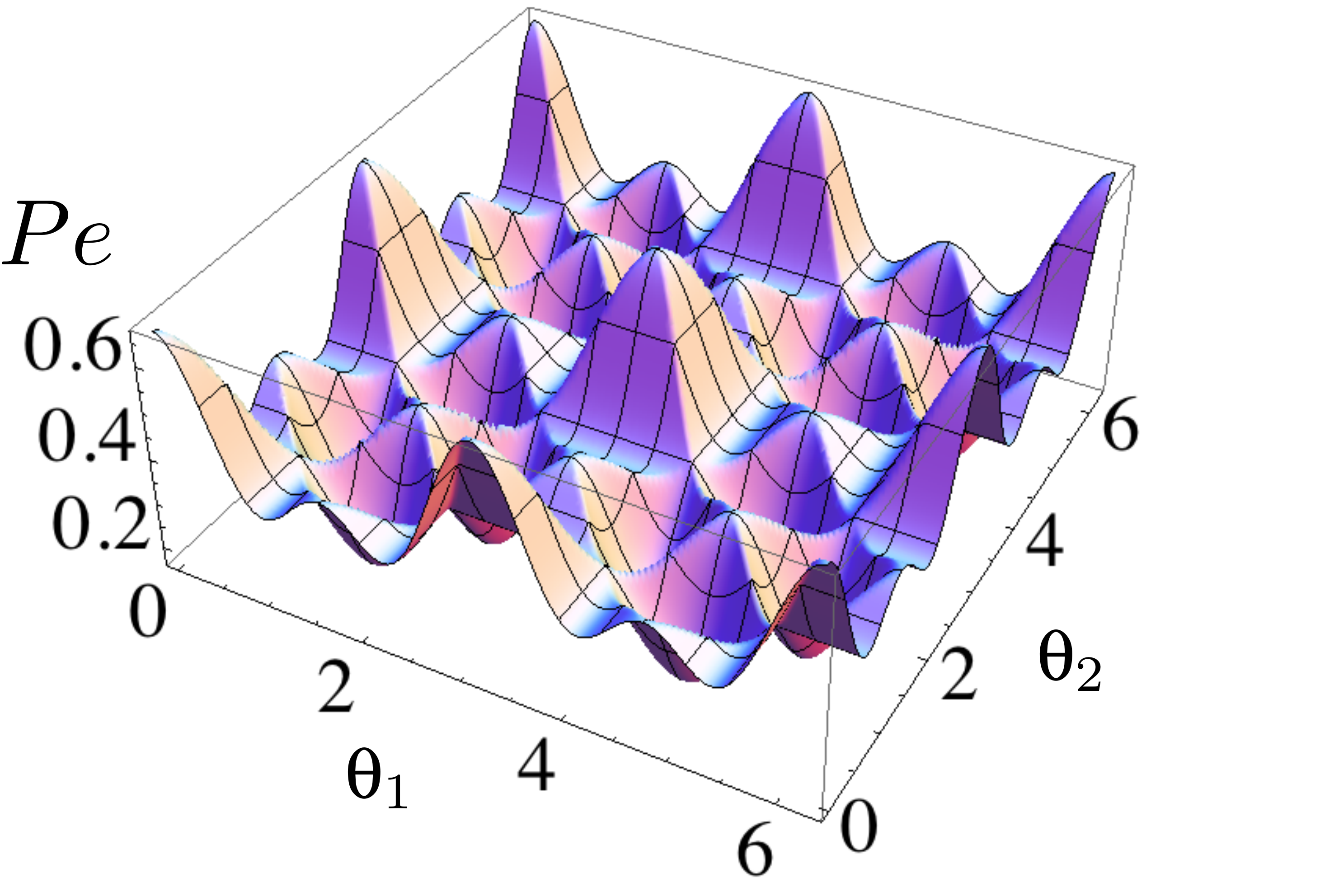}
\caption{Error probability vs phase shifts $\theta_1$ and $\theta_2$ (in radians) for optical phase discrimination between three phase shifts $(0,\theta_1,\theta_2)$.  MZI input is  $\ket{2}_a\ket{2}_b = \ket{2,0}_z$.}
\label{fig:Pe_p__M3}
\end{center}
\end{figure}

We now turn to the extension of the previous problem to discriminating three phases $(0,\theta_1,\theta_2)$ --- one of them being, again, set to zero for convenience and without loss of generality. Again, the error probability ($P_e$), is a natural criterion to assess the performance of the phase discrimination. For all phases equiprobable, the error probability is,  from \eq{Pe},
\begin{align}
P_e=&\frac{1}{3}[P(0|\theta_1)+P(0|\theta_2)+P(\theta_1|0) \nonumber\\
&+P(\theta_1|\theta_2)+P(\theta_2|0)+P(\theta_2|\theta_1)]
\end{align}
\Fig{Pe_p__M3} displays the error probability for all possible combinations of $(0,\theta_1,\theta_2)$,
for a twin-photon input. 
\begin{figure*}[t!]
\begin{center}
\includegraphics[width=\textwidth]{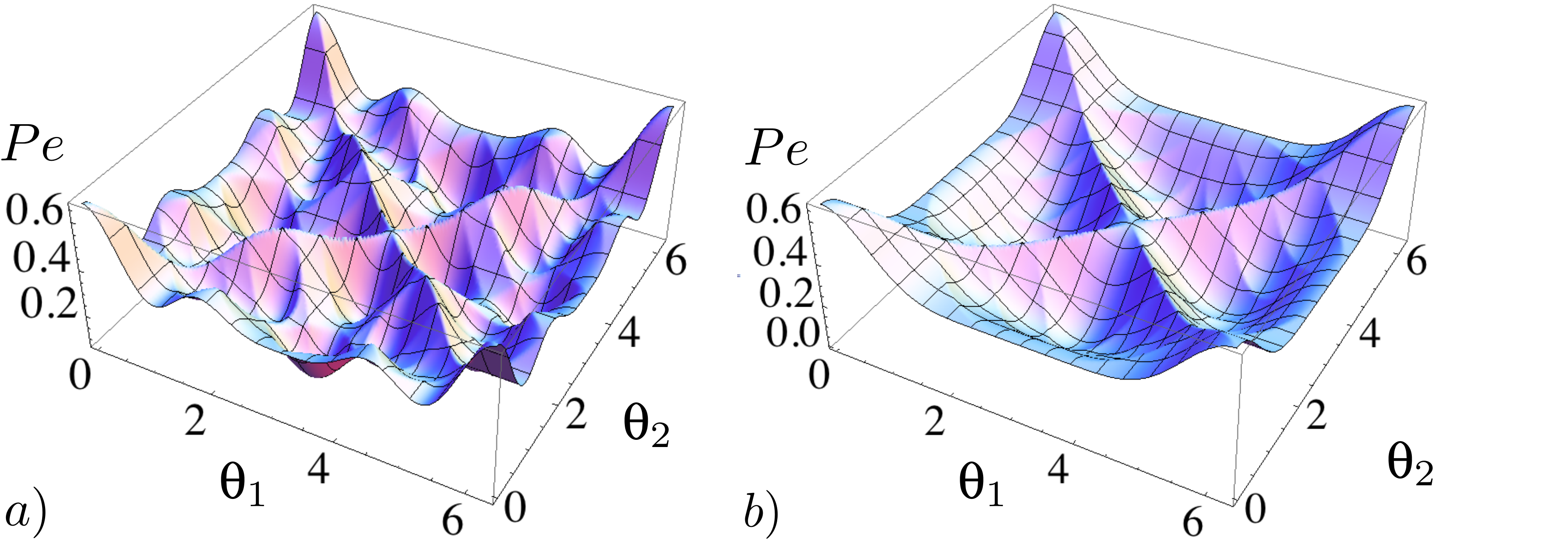}
\caption{error probability ($P_e$) vs phase shifts $\theta_1$ and $\theta_2$, for optical phase discrimination between three phase shifts $(0,\theta_1,\theta_2)$. MZI input is a), $\ket{3}_a\ket{1}_b = \ket{2,1}_z$ and  b), $\ket{4}_a\ket{0}_b = \ket{2,2}_z$}
\label{fig:Pe_p__M3jm}
\end{center}
\end{figure*}  
From the figure, we can spot the optimum phases for the problem, which label the minimums of the error probability. 

In \fig{Pe_p__M3jm}, we plot the error probability for the other two possible Fock-state inputs with $j=2$. As for the binary case, the discrimination performance depends very much on the  input state. Also, a shared trend with the binary case is that the error probability displays more oscillations, and therefore more local minima, versus the phase angle(s) as $m$ decreases, with the most favorable situation occurring for $m=0$. This case also allows discrimination of smaller phase shifts, as for the binary case.

\begin{figure}[h!]
\centerline{\includegraphics[width=\columnwidth]{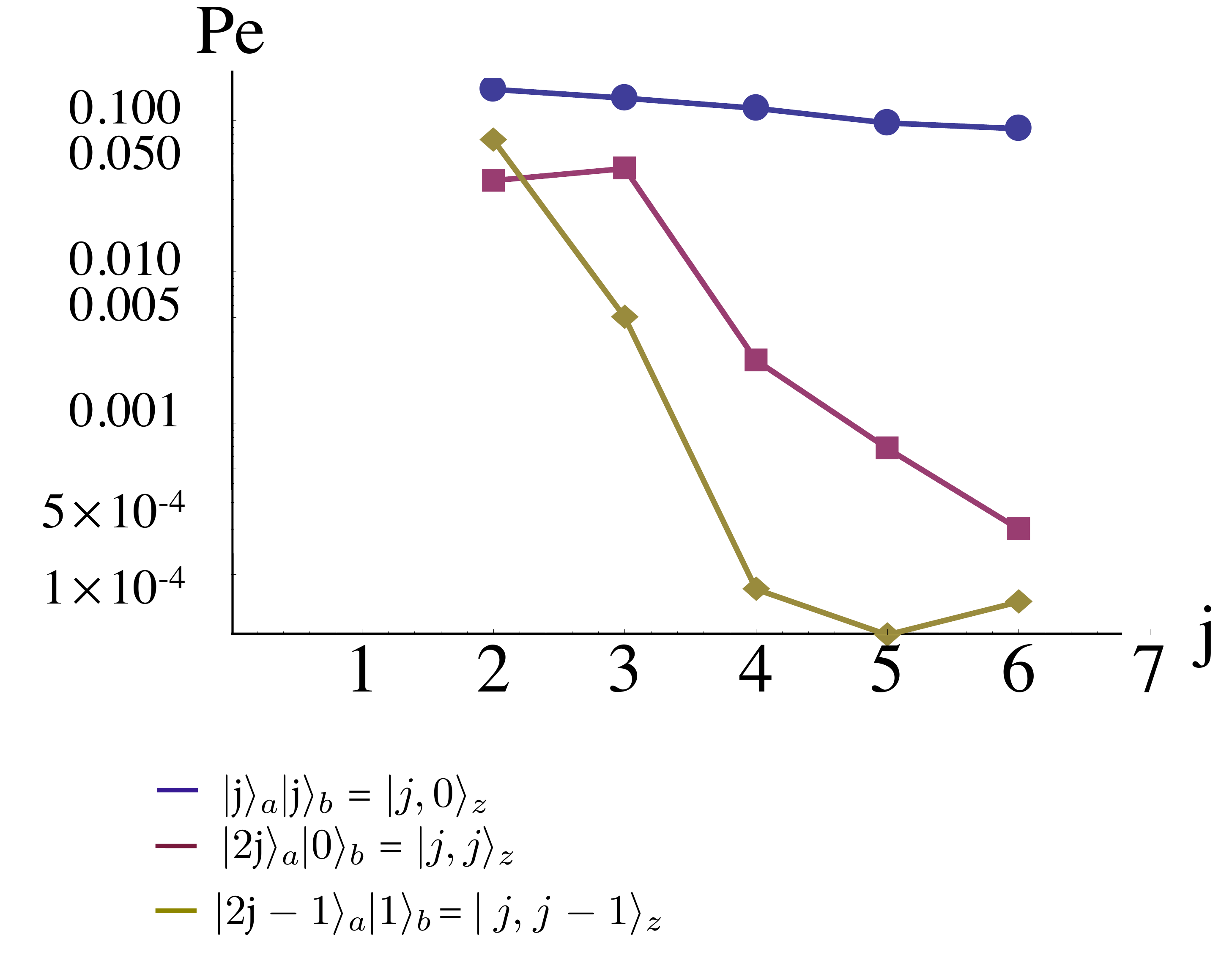}}
\vglue -.2in
\caption{Minimum error probability ($P_e$) vs N for M=3 and for all forms of input states. The effective spin $\ket{j,j-1}_z$ shows the best performance.  }
\label{fig:Pe_N__M3}
\end{figure}

We further determined optimum phase values for different input states up to $j=6$. These are listed, along with the error probability and mutual information [\eq{MI_reverse}], in \tb t.
\begin{table}[h!]
\caption{Optimal phases $\theta_{1}^{\text{opt}},\theta_{2}^{\text{opt}}$ that attain the minimum error probability $P_e^{\text{min}}$, and corresponding mutual information $I(\{\theta\};\{\hat\theta\})$ for all input states with $2\leqslant j\leqslant6$. (The maximum value of  $I(\{\theta\};\{\hat\theta\})$ here is $\log_{2}M=1.585$.)}
\vglue -.1in
\[\begin{array}{|c|cc|c|c|}
\hline
\text{Input state} & \theta_{1}^{\text{opt}}&\theta_{2}^{\text{opt}} & 10^{-3}P_{e}^{\text{min}}& I(\{\theta\};\{\hat\theta\}) \\
\hline
\ket{2\,0}_{z} = \ket2_{a}\ket2_{b}  &  \frac\pi4&\frac\pi2  & 160   &  0.93   \\
\ket{2\,1}_{z} = \ket3_{a}\ket1_{b}  &  \frac\pi2&\pi  &  160  &   0.97  \\
\ket{2\,2}_{z} = \ket4_{a}\ket0_{b}  &  \frac\pi2&\pi  &  40  &  1.35   \\
\ket{3\,0}_{z} = \ket3_{a}\ket3_{b}  &  0.67&\frac\pi2  &  140  &  1.13   \\
\ket{3\,1}_{z} = \ket4_{a}\ket2_{b}  &  \frac\pi2&\pi  &  8  &  1.50   \\
\ket{3\,2}_{z} = \ket5_{a}\ket1_{b}  &  2.32&3.16  &  5  &   1.55  \\
\ket{3\,3}_{z} = \ket6_{a}\ket0_{b}  &  \frac\pi2&\pi  &  48  &   1.30  \\
\ket{4\,0}_{z} = \ket4_{a}\ket4_{b}  &  0.55&1.2  & 120  &  1.12   \\
\ket{4\,1}_{z} = \ket5_{a}\ket3_{b}  &  1.2&\pi  &  5  &  1.52   \\
\ket{4\,2}_{z} = \ket6_{a}\ket2_{b}  &  0.6&\pi  &  2  &   1.55  \\
\ket{4\,3}_{z} = \ket7_{a}\ket1_{b}  &  0.66&\pi  &  3  &  1.56   \\
\ket{4\,4}_{z} = \ket8_{a}\ket0_{b}  &  \frac\pi2&\pi  &   3 &  1.41   \\
\ket{5\,0}_{z} = \ket5_{a}\ket5_{b}  &  0.42&\frac\pi2  &  96  &  1.20   \\
\ket{5\,1}_{z} = \ket6_{a}\ket4_{b}  &  1&\pi  &  5  &  1.54   \\
\ket{5\,2}_{z} = \ket7_{a}\ket3_{b}  &  0.48&\pi  &  3  &  1.56   \\
\ket{5\,3}_{z} = \ket8_{a}\ket2_{b}  &  0.45&\pi  &  24  &   1.40  \\
\ket{5\,4}_{z} = \ket9_{a}\ket1_{b}  &  0.64&\pi  &  0.04  &  1.58   \\
\ket{5\,5}_{z} = \ket{10}_{a}\ket0_{b}  &  \frac\pi2&\pi  & 0.68  &  1.57   \\
\ket{6\,0}_{z} = \ket6_{a}\ket6_{b}  &  1.32&2.8  &  88  &  1.20   \\
\ket{6\,1}_{z} = \ket7_{a}\ket5_{b}  &  1.32&3.16  &  4  &   1.54  \\
\ket{6\,2}_{z} = \ket8_{a}\ket4_{b}  &  0.38&\pi  &  2  &  1.56   \\
\ket{6\,3}_{z} = \ket9_{a}\ket3_{b}  &  0.42&\pi  &  0.07  &  1.58   \\
\ket{6\,4}_{z} = \ket{10}_{a}\ket2_{b}  &  2.68&\pi  &  0.23  &   1.58  \\
\ket{6\,5}_{z} = \ket{11}_{a}\ket1_{b}  &  2.56&\pi  &  0.066  &   1.58  \\
\ket{6\,6}_{z} = \ket{12}_{a}\ket0_{b}  &  1.56&\pi  &  0.2  &   1.58 \\
\hline
\end{array}
\]
\vglue -.1in
\label{tab:t}
\end{table}
Ternary discrimination does differ from binary discrimination:  while the trend that balanced inputs allow the discrimination of smaller phases as $j$ increases is confirmed by examination of all results for $m=0$, one can notice also that the error probability tends to be higher than for other cases, even though it does decrease with increasing $j$. Mutual information follows the same trend, being lowest for $m=0$ and increasing with $j$. In fact, the $m=j-1$ input offers minimal error probability and maximal mutual information for each given value of $j$, as plotted in \fig{Pe_N__M3}. This is better performance if the magnitude of the phases to be resolved isn't important. 

The problem of discriminating between more than three phases grows exponentially in size, 
and low error probabilities appear elusive for FSI with $M$$>$3.

\section{Quantum reading}\label{sec:qr}

We now examine an application of binary phase discrimination, which is the reading of digital classical memory using quantum light~\cite{pirandola_2011,Nair_2011}. 

We already know that FSI enables binary discrimination of smaller phase shifts than any other method, and also yields zero error probability at the few photon level, beating the best classical protocols. 

Here, we  compare the channel capacity and photon information efficiency (PIE) of quantum optical reading to that of classical reading and show that, whereas standard optical drives using a laser probe and direct detection are limited to a PIE of 0.5 bit of information per transmitted photon,  FSI can reach 1 bit of information per transmitted photon for binary phase discrimination, and $\log_2(3) \simeq 1.6$ bit of information per 3-phase pixel in ternary discrimination using as few as two photons.  

The classical capacity of optical reading is the amount of bits of information that can be reliably encoded and read per pixel and is equivalent to the maximum attainable mutual information  between the applied phase shifts $\theta$ and the measured phase shifts $\hat{\theta}$  for each pixel (see \app{it}),
\begin{align}
C(n_s)= \max \   I(\theta;\hat{\theta}),
\label{eq:C_optical_reading}
\end{align}
where $n_{s}$ is the average number of signal photons in the reading probe, here the interferometer arm that contains the phase shift. It is straightforward to show that
\begin{align}
{n_s}&=\bra{n_a}\bra{n_b} U_{BS}^{\dag}\, \ad a\, U_{BS}^{\dag} \ket{n_a}\ket{n_b} 
=j.
\end{align}
The PIE is then the number of bits read per signal photons:
\begin{equation}
PIE=\frac{C(n_s)}{n_s}.
\label{eq:PIE_optical_reading}
\end{equation}

Here we consider a binary phase shift keyed (BPSK) phase encoding (different from the AM scheme used in optical disks) and apply binary phase discrimination by FSI to optical reading.

\begin{figure*}[t]
\begin{center}
\includegraphics[width=.8\textwidth]{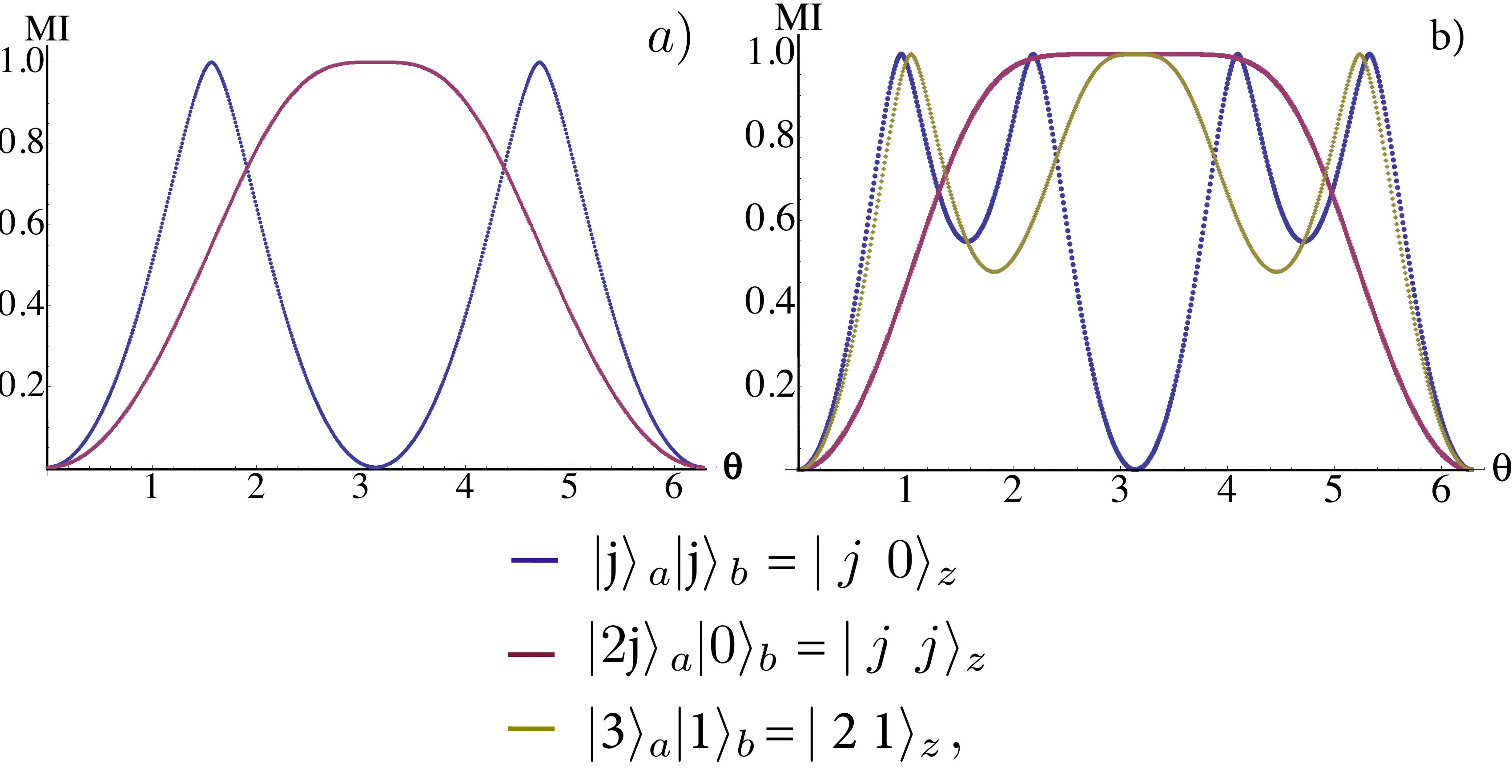}
\caption{Mutual information of optical reading. The binary information is encoded in optical phase shifts $(0,\theta)$; a), $n_{s}=j=1$; b), $n_{s}=j=2$.}
\label{fig:MI_p__M2}
\end{center}
\end{figure*}

The mutual information (MI) for FSI with $n_{s}=1,2$ photons is plotted in \fig{MI_p__M2}.  
From these results, we can determine the phases that maximize the MI. Consistent with error probability studies in the previous sections, we find that the $m=0$ states yield maximum MI for the smallest minimum phase. In the following, we use the MI at these optimum phases to calculate the channel capacity and PIE. 

\begin{figure}[h!]
\begin{center}
\centerline{\includegraphics[width=\columnwidth]{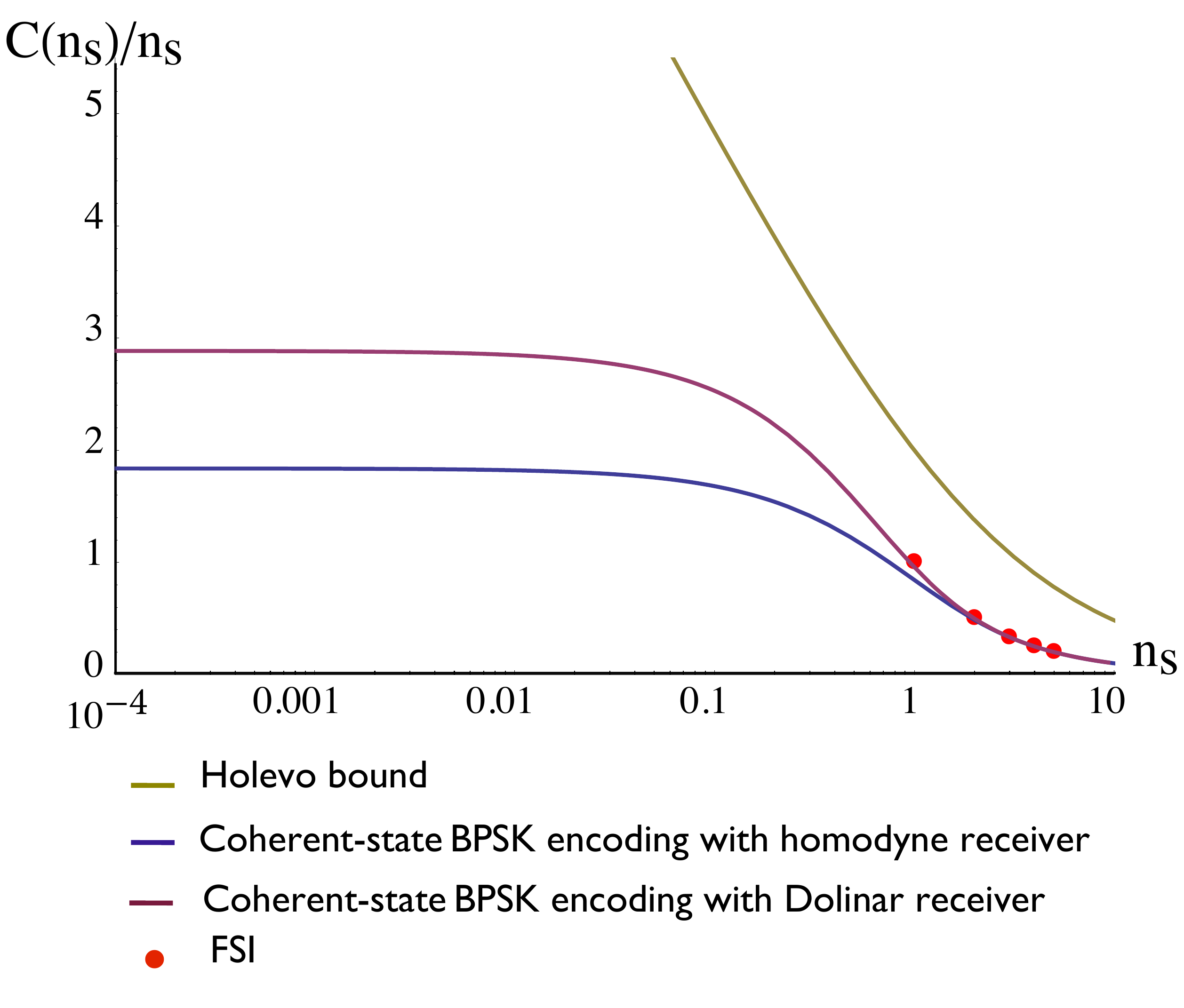}}
\caption{ Photon information efficiency versus $n_s$ for various interferometry schemes. }
\label{fig:PIEvsN_M2}
\end{center}
\end{figure}

The classical capacity of a quantum channel is limited by the Holevo bound, which imposes an upper limit on the continued reliable rate of reading classical information from a quantum channel \cite{Holevo2_1996,Holevo}. For the lossless phase only encoding the Holevo capacity can be written as \cite{Saikat_2013}
\begin{align}
C(n_s)=(1 + n_s) \log_2(1 + n_s)-n_s \log_2(n_s).
\end{align}
In order to reach the Holevo capacity bound, one needs to use an optimum probe state as well as an optimum receiver design.  However,  no feasible experimental scheme is known to achieve this limit. As for binary phase discrimination, we compare FSI performance with coherent-state encoding with homodyne, heterodyne, or Dolinar receivers. For these binary channels the  capacity is given by
 \begin{align}
&C(n_s)=1+P_e \log_2(P_e) + (1 - P_e)\log_2(1 - P_e)
\end{align}
where $P_e$ for homodyne and Dolinar  is given by \eqs{PeHomo}{PeDolinar}.   In \fig{PIEvsN_M2}, we plot the PIE versus $n_{s}$ for these different approaches.
As can be seen from the figure, the PIE of FSI slightly outperforms the Dolinar receiver for $n_{s}=1$.

\begin{figure}[h!]
\begin{center}
\centerline{\includegraphics[width=.8\columnwidth]{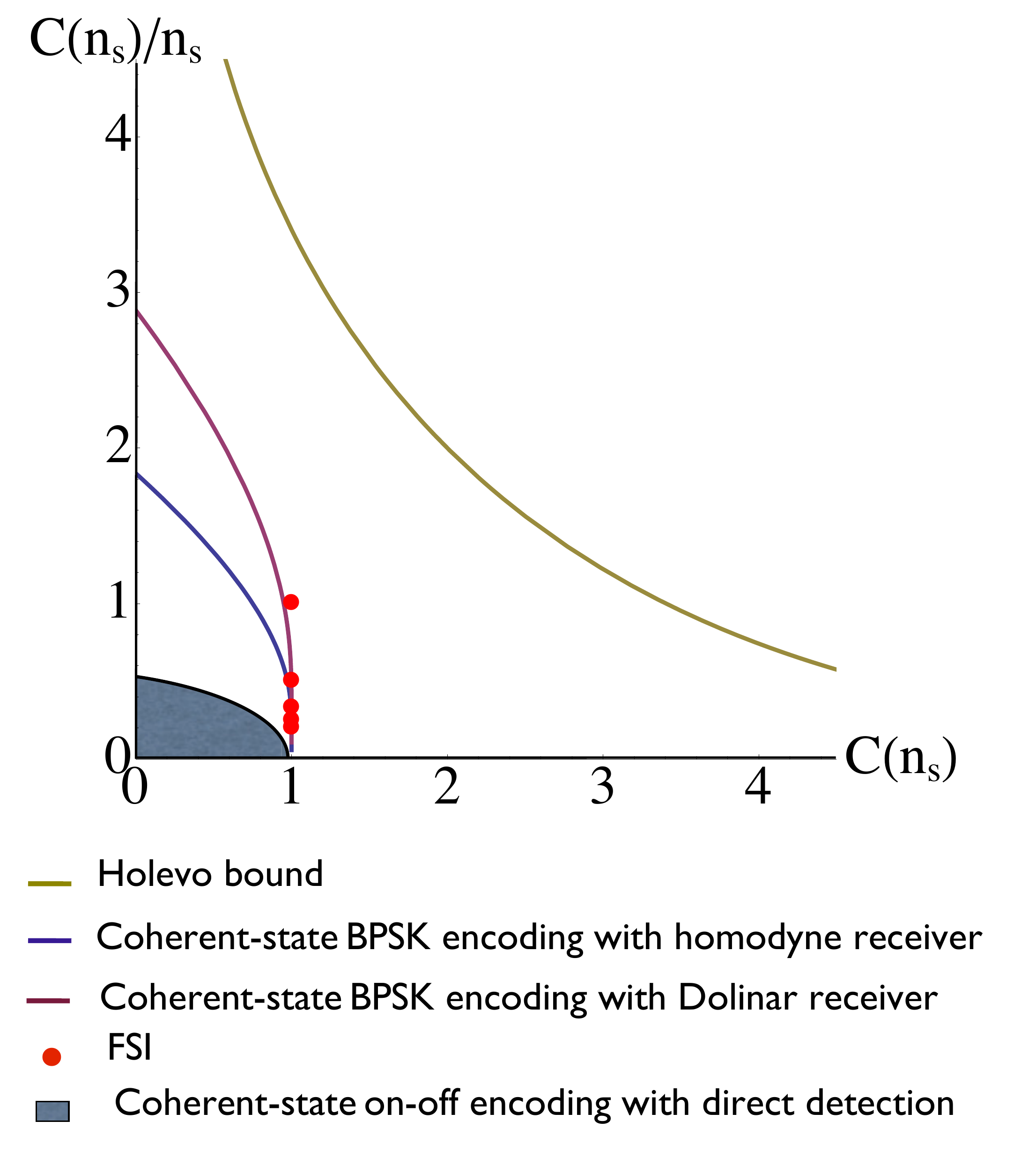}}
\caption{ Photon information efficiency (bits per photon) vs the encoding efficiency (bits encoded per pixel) for various input states and receivers.}
\label{fig:PIEvsMI_M=2}
\end{center}
\end{figure}

In \fig{PIEvsMI_M=2}, we plot the PIE versus  the channel capacity for the same BPSK receivers and also add the coherent-state on-off encoding, which is the currently used (AM) technology for reading optical disks. 
Again, FSI outperforms all other receivers.

Finally, we examine the use of ternary phase shifted keyed (TPSK) modulation, where information is encoded as three distinct phases shifts $(0,\theta_1,\theta_2)$ in each memory pixel for  optical reading, which could result in denser encoding. From our study of ternary phase discrimination in this paper, we already know the following: while an $m=0$ input will resolve smaller phases, it also does that at higher probability of error than other, ``less quantum'' inputs such as $m=j$. The latter, however, gives lower error probability for larger phase shifts, with $m=j-1$ being optimum. The results of \tb t also show that this is true for MI, which is higher, near the maximum of $\log_2(3) \simeq1.6$ for $m=j,j-1$ compared to $m=0$. Based on this we can expect a PIE of 0.8 bit/photon for $n_{s}=2$ for the $\ket{j\, j}_{z}$ input, for a choice of phases such as ($0$, $\pi/2$, $\pi$). 
  
\section{Conclusion}

We have studied Fock-state interferometry, whose implementation is now within reach with the coming of age of on-demand single-photon sources and photon-number-resolving detectors. We showed that FSI is well suited to the particular task of phase discrimination, rather than that of phase estimation. To begin with, FSI with twin inputs ($m=0$) can discriminate smaller phases  than other methods; the phase magnitude also decreases with the photon number more rapidly than $j^{-3/4}$ in the domain where our study was conducted (where 2$j$ is the total photon number). For binary discrimination, FSI can reach zero error probability; it outperforms all previously known coherent-state-based approaches, including the Dolinar receiver, at low photon levels, and matches the performance of the specifically optimized phase-eigenstate interferometry. In the more exotic case of ternary phase distribution, FSI still provides a smaller phase shift discrimination (for $m=0$ inputs), as well as very low error probability and near maximal mutual information (for $m=j-1, j$). We finally examined the application of phase discrimination to phase-shift-keyed optical reading and confirmed that the photon information efficiency of FSI exceeds all previously known experimentally feasible approaches, as well as the industry standard on-off AM reading.

\begin{acknowledgments}
This work was funded by Raytheon-BBN, under the U.S. Defense Advanced Research Projects Agency ``Information in a Photon'' program, and by  U.S. National Science Foundation grants PHY-0855632, PHY-1206029, and PHY-1708023.
\end{acknowledgments}

\appendix


\section{The Schwinger Representation}\label{app:sr}

The Schwinger representation \cite{Schwinger1965} introduces a mathematical description of passive lossless four-port optical devices based on rotations in an abstract 3D space. The application of the Schwinger representation to the analysis of optical interferometers was first demonstrated by Yurke et al.~\cite{Yurke1986a}, after W\'odkiewicz and Eberly presented the relevance to optics of the SU(2) and SU(1,1) groups~\cite{Wodkiewicz1985}. 

Any linear passive lossless optical device with two input and two output ports can be described by a $2 \times 2$  SU(2) matrix 
\begin{equation}
U=\begin{pmatrix}
\cos {\frac{\beta}{2}} e^{i(\alpha + \gamma)/2}  & \sin\frac{\beta}{2} e^{i(\alpha - \gamma)/2}\\
-\sin \frac{\beta}{2}e^{-i(\alpha + \gamma)/2} & \cos\frac{\beta}{2} e^{-i(\alpha + \gamma)/2}
\end{pmatrix}
\label{eq:SU2}
\end{equation}
where $\alpha$, $\beta$ and $\gamma$ are the Euler angles. $U$ operates on the two-dimensional vector $(a,b)^T$, whose components are the annihilation operators for the two input fields at each port of the system.
 
The homomorphism from SU(2) to the rotation group in three dimensions, SO(3), allows us to visualize the action of two-mode optical devices, such as beam splitters and phase shifters, as rotations in 3D space. The general rotation in \eq{SU2} is mathematically equivalent to the rotation of the the following tridimensional vector $\vec J$ in 3D space:
\begin{equation}
J=\begin{pmatrix} J_x \\ J_y \\ J_z \end{pmatrix}=\frac{1}{2} \begin {pmatrix} a^{\dag}b+b^{\dag}a \\ -i(a^{\dag}b-b^{\dag}a)\\ a^{\dag}a-b^{\dag}b \end{pmatrix}.
\label{eq:Jab}
\end{equation}
Components $J_x$, $J_y$ and $J_z$ follow the canonical commutation relations for quantum angular momentum operators
\begin{equation}
[  J_k, J_l ]= i \hbar  \varepsilon_{klm} \,J_m 
\label{eq:commutationJ}
\end{equation}
where $k,l,m \in \{x,y,z\} $, and $\varepsilon_{klm}$ is the Levi-Civita symbol. So $\vec J$ can be deemed  a quantum angular momentum, or effective spin.

The magnitude of the angular momentum $J^2$ is
\begin{align}
J^2&=J_x^2+J_y^2+J_z^2=\frac{a^{\dag}a+b^{\dag}b}{2}\left(\frac{a^{\dag}a+b^{\dag}b}{2}+1\right)  \\
J^{2}&=\frac{N}{2}\left(\frac{N}{2}+1\right)
\end{align}
where 
\begin{equation}
N=N_a+N_b=a^{\dag}a+b^{\dag}b
\label{eq:Nt}
\end{equation}
is the total photon number operator.

Fock states $\ket{n_a}\ket{n_b}$ are therefore also eigenstates of  $J^{2}$ and $J_z$,  
 \begin{equation}
\ket{ j\,\mu}_{z}=\ket{n_a}_{a}\ket{n_b}_{b}
\end{equation}
with respective eigenvalues $j(j+1)$ and $\mu$ given by the total photon number and the photon number difference
\begin{align}
j&=\frac{n_a+n_b}{2}\\
\mu&=\frac{n_a-n_b}{2}
\end{align}
As an example, input state of the interferometer with $2j$ photons in mode $a$ and vacuum in mode $b$ is identical to,
$\ket{2j}_a \ket{0}_b = \ket{j\,j}_z$
 and the twin Fock state input which is required for  Holland-Burnett interferometry~\cite{Holland1993} is
$\ket{j}_a \ket{j}_b =  \ket{j\,0}_z$.


A unitary operation on the quantum fields $a$ and $b$ can be viewed as the SO(3) rotation of the corresponding spin $\vec J$,  \eq{Jab}. Any  rotation of  spin $\vec J$ can be described with the 3  Euler rotations:
\begin{align}
&\vec J^\text{out}=e^{i \alpha J_z} \ e^{i \beta J_y} \ e^{i \gamma J_z} \  \vec J^\text{in} \  e^{-i \gamma J_z} \  \  e^{-i \beta J_y} \ e^{-i \alpha J_z}  \\
&\ket{\psi}_{out}=e^{i \alpha J_z} \ e^{i \beta J_y} \ e^{i \gamma J_z} \ket{\psi}_{in}
\end{align}
respectively in the Heisenberg and Schrodinger pictures.
In the Schwinger representation this arbitrary  tridimensional rotation of the effective spin $\vec J^\text{in}$ is equivalent to the Euler angle parametrization of the SU(2) rotation of the two modes $a$ and $b$ basis,  \eq{SU2}. The  SO(3)  Euler  matrix is 
 \begin{equation}
\begin{pmatrix}
c_{\alpha}c_{\beta}c_{\gamma}-s_{\alpha}s_{\gamma}& -c_{\gamma}s_{\alpha}-c_{\alpha} c_{\beta}s_{\gamma}& c_{\alpha}s_{\beta}\\
c_{\alpha} s_{\gamma}+c_{\beta}c_{\gamma}s_{\alpha}& c_{\alpha}c_{\gamma}-c_{\beta}s_{\alpha}s_{\gamma}& s_{\alpha}s_{\beta}\\
-c_{\gamma}s_{\beta}& s_{\beta}s_{\gamma}&c_{\beta}\\
\end{pmatrix}
\label{eq:Eulermatrix}
\end{equation}
 where $c_{(\alpha / \beta,\gamma)}$=$\cos{(\alpha/ \beta , \gamma)}$, $s_{(\alpha / \beta,\gamma)}$=$\sin{(\alpha/ \beta , \gamma)}$. 
The  $SO(3)$ matrix for the beam splitter with Fresnel coefficients $\rho= \cos{\phi/2}$ and $\tau= \sin{\phi/2}$ is~\cite{Yurke1986a}
 \begin{equation}
 \begin{pmatrix}
 J_x^\text{out}\\  J_y^\text{out}\\  J_z^\text{out}
 \end{pmatrix}=
\begin{pmatrix}
1&0&0\\
0& \cos{\phi}&\sin{\phi}\\
0&-\sin{\phi}& \cos{\phi}
\end{pmatrix}
 \begin{pmatrix}
 J_x^\text{in}\\  J_y^\text{in}\\  J_z^\text{in}\\
 \end{pmatrix}
\label{eq:SO3BS}
\end{equation}
The effect of the beam splitter on modes a and b is a rotation of effective spin $\vec J$ by $(-\phi)$ around the $x$ axis. The special case of the  $50/50$ beam splitter with  $\rho= \tau=\frac{1}{\sqrt{2}}$, $\phi=\pi/2$ corresponds to a $(-\pi/2)$ rotation around the $x$ axis:
\begin{align}
\vec J^\text{out}= e^{i \pi /2\,J_{x}} \ \vec J^\text{in}  \ e^{-i \pi /2\,J_{x}}\\
\ket{\psi}_{out}=e^{i \pi /2 \color{red}{J_x}} \ket{\psi}_{out}. 
\end{align}
The  SO(3)  matrix for the phase shift $\theta$ between the arms of an interferometer is
 \begin{equation}
 \begin{pmatrix}
 J_x^\text{out}\\  J_y^\text{out}\\  J_z^\text{out}
 \end{pmatrix}=
\begin{pmatrix}
\cos{\theta}&-\sin{\theta}&0\\
\sin{\theta}& \cos{\theta}&0\\
0&0&1
\end{pmatrix}
 \begin{pmatrix}
 J_x^\text{in}\\  J_y^\text{in}\\  J_z^\text{in}
 \end{pmatrix}
\label{eq:SO3Phase}
\end{equation}
which is a  rotation of effective spin around $z$ by $\theta$.
\begin{align}
\vec J^\text{out}= e^{i \theta \,J_{z}} \ \vec J^\text{in}  \ e^{- i \theta  J_{z}}\\
\ket{\psi}_{out}=e^{i \theta \color{red}{J_z}} \ket{\psi}_{out}. 
\end{align}
The Mach-Zehnder Interferometer (MZI)  consists of two $50/50$ beam splitters, and a phase shifter. So, the effect of MZI is equivalent to a $(-\pi/2)$ rotation around $x$ axis, a  $\theta$ rotation around $z$,  and another $\pi /2$ rotation around $x$, which yields a  $\theta$ rotation around $y$
\begin{align}
\vec J^\text{out}&=   e^{i \theta J_y}\ \vec J^\text{in} \  e^{-i \theta J_z\color{red}{y} } \\
\ket{\psi_{out}}& =e^{i \theta J_y} \ket{\psi_{in}}   
\end{align}
So the effect of MZI  is equivalent to a single rotation of effective spin by $\theta$ around the $y$ axis.
We are interested on the effect of MZI on Fock states, the eigenstates of effective spin $\vec J$, $\ket{j,\mu}$.
The probability function $P(\mu',\mu|\theta, j)$ for the input spin $ \ket{j,\mu}$ to be measured after the interferometer as $\ket{j,\mu'}$ for fixed $\theta$ and $J$ (the total photon number) can be described as 
a rotation matrix, which is a square matrix of dimension $2j + 1$ with general element
\begin{align}
P(\mu',\mu|\theta, j)&=|\langle j,\mu'|\psi_{out} \rangle|^2\\
&=|\langle j,\mu'|e^{i \theta J_y}|j \mu\rangle _{z}|^2\\
&=d^j_{\mu' , \mu}(\theta)^2 
\label{eq:dmatrix}
\end{align}
These rotation matrix  elements can be expressed in terms of Jacobi polynomials
\begin{align}
d^j_{\mu' , \mu}(\theta) &= \left[\frac{(j+\mu)!(j-\mu)!}{(j+\mu')!(j-\mu')!}\right]^{1/2} \(\sin{\frac{\beta}{2}}\)^{\mu-\mu'} \nonumber\\
& \times  \(\cos{\frac{\beta}{2}}\)^{\mu+\mu'}\  P_{j-\mu}^{(\mu-\mu',\mu+\mu')} (\cos{\beta})  
\end{align}

\section{Information theory}\label{app:it}

In this section we briefly recall some basic concepts of  information theory. We will use these concepts  to  study the behavior of FSI, then compare it with other schemes.

\subsection{Entropy}

For any probability distribution, we recall the definition of Shannon entropy.
Let X be a discrete random variable with the probability  function, $P(x)$, $x \in X$.\\
The Shanon entropy of the random variable $X$ is
\begin{align}
H(X)=-\sum_{x \in X} P(x) \log_2{P(x)}
\label{eq:entropy}
\end{align}
and the conditional entropy of random variables $X,Y$,  is the expected value of the entropies of the conditional distributions, averaged over the conditioning random variable,
 \begin{align}
H(Y|X) &=\sum_{x \in X} P(x)H(Y|X=x)  \\
 &= -\sum_{x \in X} P(x) \sum_{y \in Y} P(y|x) \log_{2} P(y|x)
\label{eq:cond_entropy}
\end{align}
where $P(y|x)$ is the conditional probability of measuring $y$ given that $x$ occurred.
 
\subsection{Mutual information}
  
 The mutual information (MI) is a measure of the amount of information that one random variable $X$ contains about another random variable $Y$, is equivalent to the reduction in the uncertainty of one random variable due to the knowledge of the other. For random variables $X$ and $Y$ with probability functions $P(x), x\in X$ and $P(y), y \in Y$ and the conditional entropies $H(X|Y)$ and $H(Y|X)$, the mutual information can be written as:
\begin{align}
I(X;Y)=H(X)-H(X|Y)= H(Y)-H(Y|X) 
\label{eq:mutual_information}
\end{align}
One way to study the general problem of encoding the information in $M$ optical phases, or the more specified problem of optical reading is to look at the mutual information between the applied optical phases, $\theta$ and the measured phases $\hat{\theta}$. Ideally, $\hat{\theta}$ should contain all the information about $\theta$ and the mutual information $I(\theta;\hat{\theta})$ should be equal to the amount of information in random variable $\theta$, $H(\theta)$.

 Lets assume information is encoded in $M$ optical phases with equal a priori probabilities so,
\begin{align}
\theta&=\{\theta_1, \dots ,  \theta_i , \dots , \theta_M\}\\
P(\theta)&=\left\{\frac{1}{M}, \dots , \frac{1}{M}, \dots , \frac{1}{M} \right\} \\
H(\theta)&=\log_2{M}
\end{align}
Then, one seeks to estimate the phases $\hat{\theta}$ 
\begin{align}
\hat{\theta}&=\{\hat\theta_1, \dots , \hat\theta_j, \dots , \hat\theta_M\}\\
P(\hat{\theta})&=\left\{P(\hat\theta_1), \dots , P(\hat\theta_j), \dots , P(\hat\theta_M)\right\}
\end{align}
Note that $\hat{\theta}$ is not necessarily the same random variable as $\theta$ but their similarity and the overlap in their information contents, defined as the mutual information $I(\theta;\hat{\theta})$, is a good measure of success in the phase encoding problem. 
$I(\theta;\hat{\theta})$ can be calculated as
\begin{align}
I(\theta ; \hat{\theta})&= H(\theta)-H(\theta|\hat{\theta})\\
&=H(\theta)-\sum_{j} P(\hat{\theta}_j) \sum _i P(\theta_i|\hat{\theta}_j)\log_{2}P(\theta_j|\hat{\theta}_i) 
\end{align}
$P(\hat{\theta}_j|\theta_i)$ is the probability that one can directly extract from experiment so, we employ the Bayes theorem,
 \begin{align}
P(\theta_i|\hat{\theta}_j) &= \frac{P(\hat{\theta}_j|\theta_i) P(\theta_i)}{P(\hat{\theta_j})}\\
P(\hat{\theta}_j)&=\sum_i P(\hat{\theta}_j|\theta_i) P(\theta_i)
\end{align}
and substitute $P(\theta_i|\hat{\theta}_j)$ with $P(\hat{\theta}_j|\theta_i)$ in $I(\theta ; \hat{\theta})$: 
\begin{align}
&I(\theta ; \hat{\theta})\nonumber\\
&=H(\theta)-\sum_{i,j} P(\theta_j)  \frac{P(\hat{\theta}_j|\theta_i) P(\theta_i)}{P(\hat{\theta_i})} \log_{2}\frac{P(\hat{\theta}_j|\theta_i) P(\theta_i)}{P(\hat{\theta_i})} 
\end{align}
This gives
\begin{align}
&I(\theta ; \hat{\theta})=\sum_i P(\theta_i)\log_2{P(\theta_i)} \nonumber\\
&\quad-\sum_{i,j}   P(\theta_i) P(\hat{\theta}_j|\theta_i)  \log_{2}\frac{P(\hat{\theta}_j|\theta_i) }{\sum_k P(\hat{\theta_j}|\theta_k)} \\
&=\log_2{M}-\sum_{i,j}  \frac{ P(\hat{\theta}_j|\theta_i)}{M}  \log_{2}\frac{P(\hat{\theta}_j|\theta_i) }{\sum_k P(\hat{\theta_j}|\theta_k)} 
\label{eq:MI_reverse}
\end{align}
Thus $I(\theta ; \hat{\theta})_{max}=\log_2{M}$,
 when $P(\hat{\theta}_j|\theta_i)_{i = j}=1$ and  $P(\hat{\theta}_j|\theta_i)_{i \neq j}=0$.  
  $P(\hat{\theta}_j|\theta_i)_{i \neq j}$ is the probability of having phase $\theta_i$ and estimating the wrong phase $\hat{\theta}_j$ which results in error.


\begin{thebibliography}{25}%
\makeatletter
\providecommand \@ifxundefined [1]{%
 \@ifx{#1\undefined}
}%
\providecommand \@ifnum [1]{%
 \ifnum #1\expandafter \@firstoftwo
 \else \expandafter \@secondoftwo
 \fi
}%
\providecommand \@ifx [1]{%
 \ifx #1\expandafter \@firstoftwo
 \else \expandafter \@secondoftwo
 \fi
}%
\providecommand \natexlab [1]{#1}%
\providecommand \enquote  [1]{``#1''}%
\providecommand \bibnamefont  [1]{#1}%
\providecommand \bibfnamefont [1]{#1}%
\providecommand \citenamefont [1]{#1}%
\providecommand \href@noop [0]{\@secondoftwo}%
\providecommand \href [0]{\begingroup \@sanitize@url \@href}%
\providecommand \@href[1]{\@@startlink{#1}\@@href}%
\providecommand \@@href[1]{\endgroup#1\@@endlink}%
\providecommand \@sanitize@url [0]{\catcode `\\12\catcode `\$12\catcode
  `\&12\catcode `\#12\catcode `\^12\catcode `\_12\catcode `\%12\relax}%
\providecommand \@@startlink[1]{}%
\providecommand \@@endlink[0]{}%
\providecommand \url  [0]{\begingroup\@sanitize@url \@url }%
\providecommand \@url [1]{\endgroup\@href {#1}{\urlprefix }}%
\providecommand \urlprefix  [0]{URL }%
\providecommand \Eprint [0]{\href }%
\providecommand \doibase [0]{https://doi.org/}%
\providecommand \selectlanguage [0]{\@gobble}%
\providecommand \bibinfo  [0]{\@secondoftwo}%
\providecommand \bibfield  [0]{\@secondoftwo}%
\providecommand \translation [1]{[#1]}%
\providecommand \BibitemOpen [0]{}%
\providecommand \bibitemStop [0]{}%
\providecommand \bibitemNoStop [0]{.\EOS\space}%
\providecommand \EOS [0]{\spacefactor3000\relax}%
\providecommand \BibitemShut  [1]{\csname bibitem#1\endcsname}%
\let\auto@bib@innerbib\@empty
\bibitem [{\citenamefont {Caves}(1980)}]{Caves1980}%
  \BibitemOpen
  \bibfield  {author} {\bibinfo {author} {\bibfnamefont {C.~M.}\ \bibnamefont
  {Caves}},\ }\bibfield  {title} {\bibinfo {title} {Quantum-mechanical
  radiation-pressure fluctuations in an interferometer},\ }\href@noop {}
  {\bibfield  {journal} {\bibinfo  {journal} {Phys. Rev. Lett.}\ }\textbf
  {\bibinfo {volume} {45}},\ \bibinfo {pages} {75} (\bibinfo {year}
  {1980})}\BibitemShut {NoStop}%
\bibitem [{\citenamefont {{L\'evy-Leblond}}\ and\ \citenamefont
  {Balibar}(1990)}]{JMLL}%
  \BibitemOpen
  \bibfield  {author} {\bibinfo {author} {\bibfnamefont {J.-M.}\ \bibnamefont
  {{L\'evy-Leblond}}}\ and\ \bibinfo {author} {\bibfnamefont {F.}~\bibnamefont
  {Balibar}},\ }\href@noop {} {\emph {\bibinfo {title} {Quantics: Rudiments of
  Quantum Physics}}}\ (\bibinfo  {publisher} {North-Holland},\ \bibinfo {year}
  {1990})\BibitemShut {NoStop}%
\bibitem [{\citenamefont {Caves}(1981)}]{Caves1981}%
  \BibitemOpen
  \bibfield  {author} {\bibinfo {author} {\bibfnamefont {C.~M.}\ \bibnamefont
  {Caves}},\ }\bibfield  {title} {\bibinfo {title} {Quantum-mechanical noise in
  an interferometer},\ }\href@noop {} {\bibfield  {journal} {\bibinfo
  {journal} {Phys. Rev. D}\ }\textbf {\bibinfo {volume} {23}},\ \bibinfo
  {pages} {1693} (\bibinfo {year} {1981})}\BibitemShut {NoStop}%
\bibitem [{\citenamefont {Yurke}\ \emph {et~al.}(1986)\citenamefont {Yurke},
  \citenamefont {McCall},\ and\ \citenamefont {Klauder}}]{Yurke1986a}%
  \BibitemOpen
  \bibfield  {author} {\bibinfo {author} {\bibfnamefont {B.}~\bibnamefont
  {Yurke}}, \bibinfo {author} {\bibfnamefont {S.~L.}\ \bibnamefont {McCall}},\
  and\ \bibinfo {author} {\bibfnamefont {J.~R.}\ \bibnamefont {Klauder}},\
  }\bibfield  {title} {\bibinfo {title} {{SU}(2) and {SU}(1,1)
  interferometers},\ }\href@noop {} {\bibfield  {journal} {\bibinfo  {journal}
  {Phys. Rev. A}\ }\textbf {\bibinfo {volume} {33}},\ \bibinfo {pages} {4033}
  (\bibinfo {year} {1986})}\BibitemShut {NoStop}%
\bibitem [{\citenamefont {Holland}\ and\ \citenamefont
  {Burnett}(1993)}]{Holland1993}%
  \BibitemOpen
  \bibfield  {author} {\bibinfo {author} {\bibfnamefont {M.~J.}\ \bibnamefont
  {Holland}}\ and\ \bibinfo {author} {\bibfnamefont {K.}~\bibnamefont
  {Burnett}},\ }\bibfield  {title} {\bibinfo {title} {Interferometric detection
  of optical phase shifts at the {H}eisenberg limit},\ }\href@noop {}
  {\bibfield  {journal} {\bibinfo  {journal} {Phys. Rev. Lett.}\ }\textbf
  {\bibinfo {volume} {71}},\ \bibinfo {pages} {1355} (\bibinfo {year}
  {1993})}\BibitemShut {NoStop}%
\bibitem [{\citenamefont {Kim}\ \emph {et~al.}(1998)\citenamefont {Kim},
  \citenamefont {Pfister}, \citenamefont {Holland}, \citenamefont {Noh},\ and\
  \citenamefont {Hall}}]{Kim1998}%
  \BibitemOpen
  \bibfield  {author} {\bibinfo {author} {\bibfnamefont {T.}~\bibnamefont
  {Kim}}, \bibinfo {author} {\bibfnamefont {O.}~\bibnamefont {Pfister}},
  \bibinfo {author} {\bibfnamefont {M.~J.}\ \bibnamefont {Holland}}, \bibinfo
  {author} {\bibfnamefont {J.}~\bibnamefont {Noh}},\ and\ \bibinfo {author}
  {\bibfnamefont {J.~L.}\ \bibnamefont {Hall}},\ }\bibfield  {title} {\bibinfo
  {title} {Influence of decorrelation on {H}eisenberg-limited interferometry
  using quantum correlated photons},\ }\href@noop {} {\bibfield  {journal}
  {\bibinfo  {journal} {Phys. Rev. A}\ }\textbf {\bibinfo {volume} {57}},\
  \bibinfo {pages} {4004} (\bibinfo {year} {1998})}\BibitemShut {NoStop}%
\bibitem [{\citenamefont {Luis}\ and\ \citenamefont
  {S{\'a}nchez-Soto}(2000)}]{Luis2000}%
  \BibitemOpen
  \bibfield  {author} {\bibinfo {author} {\bibfnamefont {A.}~\bibnamefont
  {Luis}}\ and\ \bibinfo {author} {\bibfnamefont {L.}~\bibnamefont
  {S{\'a}nchez-Soto}},\ }\bibfield  {title} {\bibinfo {title} {Quantum phase
  difference, phase measurements and {Stokes} operators},\ }\href@noop {}
  {\bibfield  {journal} {\bibinfo  {journal} {Prog. Opt.}\ }\textbf {\bibinfo
  {volume} {41}},\ \bibinfo {pages} {421} (\bibinfo {year} {2000})}\BibitemShut
  {NoStop}%
\bibitem [{\citenamefont {Helstrom}(1976)}]{Helstrom_1976}%
  \BibitemOpen
  \bibfield  {author} {\bibinfo {author} {\bibfnamefont {C.~W.}\ \bibnamefont
  {Helstrom}},\ }\href@noop {} {\emph {\bibinfo {title} {Quantum Detection and
  Estimation Theory}}}\ (\bibinfo  {publisher} {Mathematics in Science and
  Engineering, 123,({A}cademic {P}ress, {N}ew {Y}ork)},\ \bibinfo {year}
  {1976})\BibitemShut {NoStop}%
\bibitem [{\citenamefont {Giovannetti}\ \emph {et~al.}(2006)\citenamefont
  {Giovannetti}, \citenamefont {Lloyd},\ and\ \citenamefont
  {Maccone}}]{Giovani:2006_Qmetrology}%
  \BibitemOpen
  \bibfield  {author} {\bibinfo {author} {\bibfnamefont {V.}~\bibnamefont
  {Giovannetti}}, \bibinfo {author} {\bibfnamefont {S.}~\bibnamefont {Lloyd}},\
  and\ \bibinfo {author} {\bibfnamefont {L.}~\bibnamefont {Maccone}},\
  }\bibfield  {title} {\bibinfo {title} {Quantum metrology},\ }\href
  {https://doi.org/10.1103/PhysRevLett.96.010401} {\bibfield  {journal}
  {\bibinfo  {journal} {Phys. Rev. Lett.}\ }\textbf {\bibinfo {volume} {96}},\
  \bibinfo {pages} {010401} (\bibinfo {year} {2006})}\BibitemShut {NoStop}%
\bibitem [{\citenamefont {Escher}\ \emph {et~al.}(2011)\citenamefont {Escher},
  \citenamefont {de~Matos~Filho},\ and\ \citenamefont
  {Davidovich}}]{Escher2011}%
  \BibitemOpen
  \bibfield  {author} {\bibinfo {author} {\bibfnamefont {B.~M.}\ \bibnamefont
  {Escher}}, \bibinfo {author} {\bibfnamefont {R.~L.}\ \bibnamefont
  {de~Matos~Filho}},\ and\ \bibinfo {author} {\bibfnamefont {L.}~\bibnamefont
  {Davidovich}},\ }\bibfield  {title} {\bibinfo {title} {General framework for
  estimating the ultimate precision limit in noisy quantum-enhanced
  metrology},\ }\href {http://dx.doi.org/10.1038/nphys1958} {\bibfield
  {journal} {\bibinfo  {journal} {Nat. Phys.}\ }\textbf {\bibinfo {volume}
  {7}},\ \bibinfo {pages} {406} (\bibinfo {year} {2011})}\BibitemShut {NoStop}%
\bibitem [{\citenamefont {Thekkadath}\ \emph {et~al.}(2020)\citenamefont
  {Thekkadath}, \citenamefont {Mycroft}, \citenamefont {Bell}, \citenamefont
  {Wade}, \citenamefont {Eckstein}, \citenamefont {Phillips}, \citenamefont
  {Patel}, \citenamefont {Buraczewski}, \citenamefont {Lita}, \citenamefont
  {Gerrits}, \citenamefont {Nam}, \citenamefont {Stobi{\'n}ska}, \citenamefont
  {Lvovsky},\ and\ \citenamefont {Walmsley}}]{Thekkadath2020}%
  \BibitemOpen
  \bibfield  {author} {\bibinfo {author} {\bibfnamefont {G.~S.}\ \bibnamefont
  {Thekkadath}}, \bibinfo {author} {\bibfnamefont {M.~E.}\ \bibnamefont
  {Mycroft}}, \bibinfo {author} {\bibfnamefont {B.~A.}\ \bibnamefont {Bell}},
  \bibinfo {author} {\bibfnamefont {C.~G.}\ \bibnamefont {Wade}}, \bibinfo
  {author} {\bibfnamefont {A.}~\bibnamefont {Eckstein}}, \bibinfo {author}
  {\bibfnamefont {D.~S.}\ \bibnamefont {Phillips}}, \bibinfo {author}
  {\bibfnamefont {R.~B.}\ \bibnamefont {Patel}}, \bibinfo {author}
  {\bibfnamefont {A.}~\bibnamefont {Buraczewski}}, \bibinfo {author}
  {\bibfnamefont {A.~E.}\ \bibnamefont {Lita}}, \bibinfo {author}
  {\bibfnamefont {T.}~\bibnamefont {Gerrits}}, \bibinfo {author} {\bibfnamefont
  {S.~W.}\ \bibnamefont {Nam}}, \bibinfo {author} {\bibfnamefont
  {M.}~\bibnamefont {Stobi{\'n}ska}}, \bibinfo {author} {\bibfnamefont {A.~I.}\
  \bibnamefont {Lvovsky}},\ and\ \bibinfo {author} {\bibfnamefont {I.~A.}\
  \bibnamefont {Walmsley}},\ }\bibfield  {title} {\bibinfo {title}
  {Quantum-enhanced interferometry with large heralded photon-number states},\
  }\href {https://doi.org/10.1038/s41534-020-00320-y} {\bibfield  {journal}
  {\bibinfo  {journal} {npj Quantum Information}\ }\textbf {\bibinfo {volume}
  {6}},\ \bibinfo {pages} {89} (\bibinfo {year} {2020})}\BibitemShut {NoStop}%
\bibitem [{\citenamefont {Perarnau-Llobet}\ \emph {et~al.}(2020)\citenamefont
  {Perarnau-Llobet}, \citenamefont {Gonz{\'{a}}lez-Tudela},\ and\ \citenamefont
  {Cirac}}]{Perarnau_Llobet_2020}%
  \BibitemOpen
  \bibfield  {author} {\bibinfo {author} {\bibfnamefont {M.}~\bibnamefont
  {Perarnau-Llobet}}, \bibinfo {author} {\bibfnamefont {A.}~\bibnamefont
  {Gonz{\'{a}}lez-Tudela}},\ and\ \bibinfo {author} {\bibfnamefont {J.~I.}\
  \bibnamefont {Cirac}},\ }\bibfield  {title} {\bibinfo {title} {Multimode fock
  states with large photon number: effective descriptions and applications in
  quantum metrology},\ }\href {https://doi.org/10.1088/2058-9565/ab6ce5}
  {\bibfield  {journal} {\bibinfo  {journal} {Quantum Science and Technology}\
  }\textbf {\bibinfo {volume} {5}},\ \bibinfo {pages} {025003} (\bibinfo {year}
  {2020})}\BibitemShut {NoStop}%
\bibitem [{\citenamefont {Schwinger}()}]{Schwinger1965}%
  \BibitemOpen
  \bibfield  {author} {\bibinfo {author} {\bibfnamefont {J.}~\bibnamefont
  {Schwinger}},\ }\bibfield  {title} {\bibinfo {title} {On angular momentum},\
  }\href {http://www.osti.gov/scitech/biblio/4389568} {\bibinfo  {journal}
  {U.S. Atomic Energy Commission Report. No. NYO--3071 (1952), reprinted in
  Quantum Theory of Angular Momentum, edited by L. C. Biedenharn and H. van Dam
  (Academic Press, New York, 1965), pp. 229--279}\ }\BibitemShut {NoStop}%
\bibitem [{\citenamefont {Agrawal}(2010)}]{Agrawal_2010_opt_commmunication}%
  \BibitemOpen
\bibfield  {journal} {  }\bibfield  {author} {\bibinfo {author} {\bibfnamefont
  {G.~P.}\ \bibnamefont {Agrawal}},\ }\href@noop {} {\emph {\bibinfo {title}
  {Fiber-Optic Communication Systems}}},\ \bibinfo {edition} {4th}\ ed.\
  (\bibinfo  {publisher} {Wiley Series in Microwave and Optical Engineering,},\
  \bibinfo {address} {Wiley- Interscience, Hoboken, NJ},\ \bibinfo {year}
  {2010})\BibitemShut {NoStop}%
\bibitem [{\citenamefont {Braunstein}\ and\ \citenamefont {van
  Loock}(2005)}]{Braunstein_2005_Homodyne_det}%
  \BibitemOpen
  \bibfield  {author} {\bibinfo {author} {\bibfnamefont {S.~L.}\ \bibnamefont
  {Braunstein}}\ and\ \bibinfo {author} {\bibfnamefont {P.}~\bibnamefont {van
  Loock}},\ }\bibfield  {title} {\bibinfo {title} {Quantum information with
  continuous variables},\ }\href {https://doi.org/10.1103/RevModPhys.77.513}
  {\bibfield  {journal} {\bibinfo  {journal} {Rev. Mod. Phys.}\ }\textbf
  {\bibinfo {volume} {77}},\ \bibinfo {pages} {513} (\bibinfo {year}
  {2005})}\BibitemShut {NoStop}%
\bibitem [{\citenamefont {Olivares}\ and\ \citenamefont
  {Paris}(2004)}]{Oliviars_2004_homodyne_error}%
  \BibitemOpen
  \bibfield  {author} {\bibinfo {author} {\bibfnamefont {S.}~\bibnamefont
  {Olivares}}\ and\ \bibinfo {author} {\bibfnamefont {M.~G.~A.}\ \bibnamefont
  {Paris}},\ }\bibfield  {title} {\bibinfo {title} {Binary optical
  communication in single-mode and entangled quantum noisy channels},\
  }\href@noop {} {\bibfield  {journal} {\bibinfo  {journal} {Journal of Optics
  B: Quantum and Semiclassical Optics}\ }\textbf {\bibinfo {volume} {Opt. 6
  69}} (\bibinfo {year} {2004})}\BibitemShut {NoStop}%
\bibitem [{\citenamefont {Dolinar}(1973)}]{Dolinar1973}%
  \BibitemOpen
  \bibfield  {author} {\bibinfo {author} {\bibfnamefont {S.}~\bibnamefont
  {Dolinar}},\ }\bibfield  {title} {\bibinfo {title} {Quarterly progress
  report},\ }\href@noop {} {\bibfield  {journal} {\bibinfo  {journal} {Tech.
  Rep., RLE at MIT}\ }\textbf {\bibinfo {volume} {111}} (\bibinfo {year}
  {1973})}\BibitemShut {NoStop}%
\bibitem [{\citenamefont {Nair}\ \emph {et~al.}(2012)\citenamefont {Nair},
  \citenamefont {Yen}, \citenamefont {Guha}, \citenamefont {Shapiro},\ and\
  \citenamefont {Pirandola}}]{saikat:2012_symphasedisc}%
  \BibitemOpen
  \bibfield  {author} {\bibinfo {author} {\bibfnamefont {R.}~\bibnamefont
  {Nair}}, \bibinfo {author} {\bibfnamefont {B.~J.}\ \bibnamefont {Yen}},
  \bibinfo {author} {\bibfnamefont {S.}~\bibnamefont {Guha}}, \bibinfo {author}
  {\bibfnamefont {J.~H.}\ \bibnamefont {Shapiro}},\ and\ \bibinfo {author}
  {\bibfnamefont {S.}~\bibnamefont {Pirandola}},\ }\bibfield  {title} {\bibinfo
  {title} {Symmetric m-ary phase discrimination using quantum-optical probe
  states},\ }\href {https://doi.org/10.1103/PhysRevA.86.022306} {\bibfield
  {journal} {\bibinfo  {journal} {Phys. Rev. A}\ }\textbf {\bibinfo {volume}
  {86}},\ \bibinfo {pages} {022306} (\bibinfo {year} {2012})}\BibitemShut
  {NoStop}%
\bibitem [{\citenamefont {Pegg}\ and\ \citenamefont
  {Barnett}(1989)}]{Pegg1989}%
  \BibitemOpen
  \bibfield  {author} {\bibinfo {author} {\bibfnamefont {D.~T.}\ \bibnamefont
  {Pegg}}\ and\ \bibinfo {author} {\bibfnamefont {S.~M.}\ \bibnamefont
  {Barnett}},\ }\bibfield  {title} {\bibinfo {title} {Phase properties of the
  quantized single-mode electromagnetic field},\ }\href
  {https://doi.org/10.1103/PhysRevA.39.1665} {\bibfield  {journal} {\bibinfo
  {journal} {Phys. Rev. A}\ }\textbf {\bibinfo {volume} {39}},\ \bibinfo
  {pages} {1665} (\bibinfo {year} {1989})}\BibitemShut {NoStop}%
\bibitem [{\citenamefont {Pirandola}(2011)}]{pirandola_2011}%
  \BibitemOpen
  \bibfield  {author} {\bibinfo {author} {\bibfnamefont {S.}~\bibnamefont
  {Pirandola}},\ }\bibfield  {title} {\bibinfo {title} {Quantum reading of a
  classical digital memory},\ }\href
  {https://doi.org/10.1103/PhysRevLett.106.090504} {\bibfield  {journal}
  {\bibinfo  {journal} {Phys. Rev. Lett.}\ }\textbf {\bibinfo {volume} {106}},\
  \bibinfo {pages} {090504} (\bibinfo {year} {2011})}\BibitemShut {NoStop}%
\bibitem [{\citenamefont {Nair}(2011)}]{Nair_2011}%
  \BibitemOpen
  \bibfield  {author} {\bibinfo {author} {\bibfnamefont {R.}~\bibnamefont
  {Nair}},\ }\bibfield  {title} {\bibinfo {title} {Discriminating
  quantum-optical beam-splitter channels with number-diagonal signal states:
  Applications to quantum reading and target detection},\ }\href
  {https://doi.org/10.1103/PhysRevA.84.032312} {\bibfield  {journal} {\bibinfo
  {journal} {Phys. Rev. A}\ }\textbf {\bibinfo {volume} {84}},\ \bibinfo
  {pages} {032312} (\bibinfo {year} {2011})}\BibitemShut {NoStop}%
\bibitem [{\citenamefont {Hausladen}\ \emph {et~al.}(1996)\citenamefont
  {Hausladen}, \citenamefont {Jozsa}, \citenamefont {Schumacher}, \citenamefont
  {Westmoreland},\ and\ \citenamefont {Wootters}}]{Holevo2_1996}%
  \BibitemOpen
  \bibfield  {author} {\bibinfo {author} {\bibfnamefont {P.}~\bibnamefont
  {Hausladen}}, \bibinfo {author} {\bibfnamefont {R.}~\bibnamefont {Jozsa}},
  \bibinfo {author} {\bibfnamefont {B.}~\bibnamefont {Schumacher}}, \bibinfo
  {author} {\bibfnamefont {M.}~\bibnamefont {Westmoreland}},\ and\ \bibinfo
  {author} {\bibfnamefont {W.~K.}\ \bibnamefont {Wootters}},\ }\bibfield
  {title} {\bibinfo {title} {Classical information capacity of a quantum
  channel},\ }\href {https://doi.org/10.1103/PhysRevA.54.1869} {\bibfield
  {journal} {\bibinfo  {journal} {Phys. Rev. A}\ }\textbf {\bibinfo {volume}
  {54}},\ \bibinfo {pages} {1869} (\bibinfo {year} {1996})}\BibitemShut
  {NoStop}%
\bibitem [{\citenamefont {Holevo}(1979)}]{Holevo}%
  \BibitemOpen
  \bibfield  {author} {\bibinfo {author} {\bibfnamefont {A.~S.}\ \bibnamefont
  {Holevo}},\ }\href@noop {} {\bibfield  {journal} {\bibinfo  {journal}
  {Problemy Peredachi lnformatsii 15, 3 [Problems of lnformation Transmission
  (USSR)}\ }\textbf {\bibinfo {volume} {15}},\ \bibinfo {pages} {247} (\bibinfo
  {year} {1979})}\BibitemShut {NoStop}%
\bibitem [{\citenamefont {Guha}\ and\ \citenamefont
  {Shapiro}(2013)}]{Saikat_2013}%
  \BibitemOpen
  \bibfield  {author} {\bibinfo {author} {\bibfnamefont {S.}~\bibnamefont
  {Guha}}\ and\ \bibinfo {author} {\bibfnamefont {J.~H.}\ \bibnamefont
  {Shapiro}},\ }\bibfield  {title} {\bibinfo {title} {Reading boundless
  error-free bits using a single photon},\ }\href
  {https://doi.org/10.1103/PhysRevA.87.062306} {\bibfield  {journal} {\bibinfo
  {journal} {Phys. Rev. A}\ }\textbf {\bibinfo {volume} {87}},\ \bibinfo
  {pages} {062306} (\bibinfo {year} {2013})}\BibitemShut {NoStop}%
\bibitem [{\citenamefont {W\'odkiewicz}\ and\ \citenamefont
  {Eberly}(1985)}]{Wodkiewicz1985}%
  \BibitemOpen
  \bibfield  {author} {\bibinfo {author} {\bibfnamefont {K.}~\bibnamefont
  {W\'odkiewicz}}\ and\ \bibinfo {author} {\bibfnamefont {J.}~\bibnamefont
  {Eberly}},\ }\bibfield  {title} {\bibinfo {title} {Coherent states, squeezed
  fluctuations, and the {SU(2)} and {SU(1,1)} groups in quantum-optics
  applications},\ }\href@noop {} {\bibfield  {journal} {\bibinfo  {journal} {J.
  Opt. Soc. Am. B}\ }\textbf {\bibinfo {volume} {2}},\ \bibinfo {pages} {458}
  (\bibinfo {year} {1985})}\BibitemShut {NoStop}%
\end{thebibliography}
%

\end{document}